\documentclass[apj,tighten,twocolumn,numberedappendix,twocolappendix]{emulateapj}
\usepackage{amsmath}
\usepackage{hyperref}
\usepackage{natbib}
\usepackage{savesym}
\savesymbol{tablenum}
\usepackage{siunitx}
\restoresymbol{SIX}{tablenum}
\usepackage{verbatim}
\usepackage{xspace}

\newcommand{\appref}[1]{Appendix~\ref{app:#1}}
\newcommand{\figref}[1]{Figure~\ref{fig:#1}}
\newcommand{\tabref}[1]{Table~\ref{tab:#1}}
\newcommand{\secref}[1]{\S\ref{sec:#1}}

\newcommand{\subsecref}[1]{\S\ref{subsec:#1}}

\newcommand{\sol}{\odot}
\newcommand{\HI}{H\,{\sc i}\xspace}
\newcommand{\cmaes}{{\textsc CMAES}\xspace}
\newcommand{\galactics}{{\textsc GalactICS}\xspace}
\newcommand{\laplacesdemon}{{\textsc LaplacesDemon}\xspace}
\newcommand{\magrite}{{\textsc MagRite}\xspace}
\newcommand{\profit}{{\textsc ProFit}\xspace}
\newcommand{\R}{{\textsc R}\xspace}
\newcommand{\sersic}{{\textsc S\'{e}rsic}\xspace}
\newcommand{\kms}{$\mathrm{km s^{-1}}$}
\newcommand{\kmse}{\mathrm{km s^{-1}}}

\newcommand{\lsole}{\mathrm{L_{\sol}}}
\newcommand{\msol}{$\mathrm{M_{\sol}}$}
\newcommand{\msole}{\mathrm{M_{\sol}}}

\newcommand{\zsole}{\mathrm{Z_{\sol}}}
\newcommand{\ec}{{\cal E}}
\def\mean#1{\left< #1 \right>}

\shorttitle{Self-consistent IFS galaxy modelling}
\shortauthors{Taranu et al.}

\begin{document}

\title{Self-consistent bulge/disk/halo galaxy dynamical modeling using integral field kinematics}

\begin{abstract}
We introduce a method for modeling disk galaxies designed to take full advantage of data from integral field spectroscopy (IFS). The method fits equilibrium models to simultaneously reproduce the surface brightness, rotation and velocity dispersion profiles of a galaxy. The models are fully self-consistent 6D distribution functions for a galaxy with a Sersic-profile stellar bulge, exponential disk and parametric dark matter halo, generated by an updated version of GalactICS. By creating realistic flux-weighted maps of the kinematic moments (flux, mean velocity and dispersion), we simultaneously fit photometric and spectroscopic data using both maximum-likelihood and Bayesian (MCMC) techniques. We apply the method to a GAMA spiral galaxy (G79635) with kinematics from the SAMI Galaxy Survey and deep $g$- and $r$-band photometry from the VST-KiDS survey, comparing parameter constraints with those from traditional 2D bulge-disk decomposition. Our method returns broadly consistent results for shared parameters, while constraining the mass-to-light ratios of stellar components and reproducing the \HI-inferred circular velocity well beyond the limits of the SAMI data. While the method is tailored for fitting integral field kinematic data, it can use other dynamical constraints like central fibre dispersions and \HI circular velocities, and is well-suited for modelling galaxies with a combination of deep imaging and \HI and/or optical spectra (resolved or otherwise). Our implementation (MagRite) is computationally efficient and can generate well-resolved models and kinematic maps in under a minute on modern processors.
\end{abstract}

\author{
D.~S. Taranu$^{1,2}$, 
D. Obreschkow$^{1,2}$,
J.~J. Dubinski$^{3}$,
L.~M.~R. Fogarty$^{2,4}$,
J. van de Sande$^{4}$,
B. Catinella$^{1}$,
L. Cortese$^{1}$,
A. Moffett$^{1}$,
A.~S.~G. Robotham$^{1}$,
J.~T. Allen$^{2,4}$,
J. Bland-Hawthorn$^{4}$,
J.~J. Bryant$^{2,4,5}$,
M. Colless$^{2,6}$,
S.~M. Croom$^{2,4}$,
F. D'Eugenio$^{6}$,
R.~L. Davies$^{7}$,
M.~J. Drinkwater$^{2,8}$,
S.~P. Driver$^{1}$,
M. Goodwin$^{5}$,
I.~S. Konstantopoulos$^{9}$,
J.~S. Lawrence$^{5}$,
\'A.~R. L\'opez-S\'anchez$^{5,10}$,
N.~P.~F. Lorente$^{5}$,
A.~M. Medling$^{6,11,12}$,
J.~R. Mould$^{13}$,
M.~S. Owers$^{5,10}$,
C. Power$^{1,2}$,
S.~N. Richards$^{2,4,5}$,
C. Tonini$^{14}$
}
\altaffiltext{1}{International Centre for Radio Astronomy Research, The University of Western Australia, 35 Stirling Highway, Crawley, WA, 6009, Australia}
\altaffiltext{2}{ARC Centre of Excellence for All-sky Astrophysics (CAASTRO)}
\altaffiltext{3}{Department of Astronomy and Astrophysics, University of Toronto, 50 St. George Street, Toronto, ON, Canada, M5S 3H4}
\altaffiltext{4}{Sydney Institute for Astronomy, School of Physics, A28, The University of Sydney, NSW, 2006, Australia}
\altaffiltext{5}{Australian Astronomical Observatory, PO Box 915, North Ryde, NSW, 1670, Australia}
\altaffiltext{6}{Research School of Astronomy and Astrophysics, Australian National University, Canberra, ACT, 2611, Australia}
\altaffiltext{7}{Astrophysics,  Department of Physics, University of Oxford, Denys Wilkinson Building, Keble Rd., Oxford, OX1 3RH, UK}
\altaffiltext{8}{School of Mathematics and Physics, University of Queensland, QLD, 4072, Australia}
\altaffiltext{9}{Envizi Suite 213, National Innovation Centre, Australian Technology Park, 4 Cornwallis Street, Eveleigh NSW 2015, Australia}
\altaffiltext{10}{Department of Physics and Astronomy, Macquarie University, NSW, 2109, Australia}
\altaffiltext{11}{Cahill Center for Astronomy and Astrophysics California Institute of Technology, MS 249-17, Pasadena, CA, 91125, USA}
\altaffiltext{12}{Hubble Fellow}
\altaffiltext{13}{Swinburne University, Hawthorn, VIC, 3122, Australia}
\altaffiltext{14}{Melbourne University, School of Physics, Parkville, VIC, 3010, Australia}


\section{Introduction}
\label{sec:introduction}

The physical properties of nearby spiral galaxies are typically derived by fitting a number of distinct components to broadband images, either using azimuthally-averaged 1D profiles or directly in 2D \citep[e.g.][]{PenHoImp02,SanFerMac16,JohHaeAra17}. For large surveys, common models are single \citet{Ser68} profile fits or two-component bulge-disk decompositions using an exponential disk and a \cite{deV59} or \sersic profile bulge \citep{SimMenPat11} - appropriate parameterizations for galaxy disks and ``classical'' dispersion-supported bulges \citep{Gad09}, respectively. Additional features like bars, spiral arms and dust are usually only modelled for well-resolved nearby galaxies.

Photometric bulge-disk decomposition has several major drawbacks. Firstly, the best-fit 2D model may be impossible to reproduce with more realistic 3D density profiles or a 6D phase space distribution function (DF) - a serious concern, since most nearby galaxies are dynamically relaxed systems close to virial equilibrium. 2D models may be unable to produce a stable equilibrium system, or require an unrealistic dark matter halo density profile to reproduce the rotation curve. Therefore, it is desirable that fitting methods exclude parameter combinations that cannot create stable equilibrium models consistent with the galaxy's dynamics.

Bulge-disk decompositions can also produce ambiguous results. For fits with an exponential and a \sersic profile, it is often assumed that the exponential component is a disk, whereas the \sersic component is a bulge; however, the bulge can have a best-fit \sersic index $n_{s}\approx1$, leaving only the size and ellipticity to distinguish it from the disk. Furthermore, bulges are typically centrally concentrated and compact, and therefore difficult to resolve beyond $z>0.05$ with seeing-limited imaging \citep{KelDriRob14}.

Kinematic data can break these degeneracies, even if spatially unresolved. A single-fibre central velocity dispersion can be used to infer the presence of a ``classical'', dispersion-supported bulge. Similarly, an unresolved \HI 21cm spectrum exhibiting a ``double-horned'' profile traces the orbital velocity of the neutral hydrogen gas, constraining the circular velocity at a large physical radius and (in principle) the combination of the rotation curve and \HI surface density profile.

\begin{figure*}
\includegraphics[width=\textwidth]{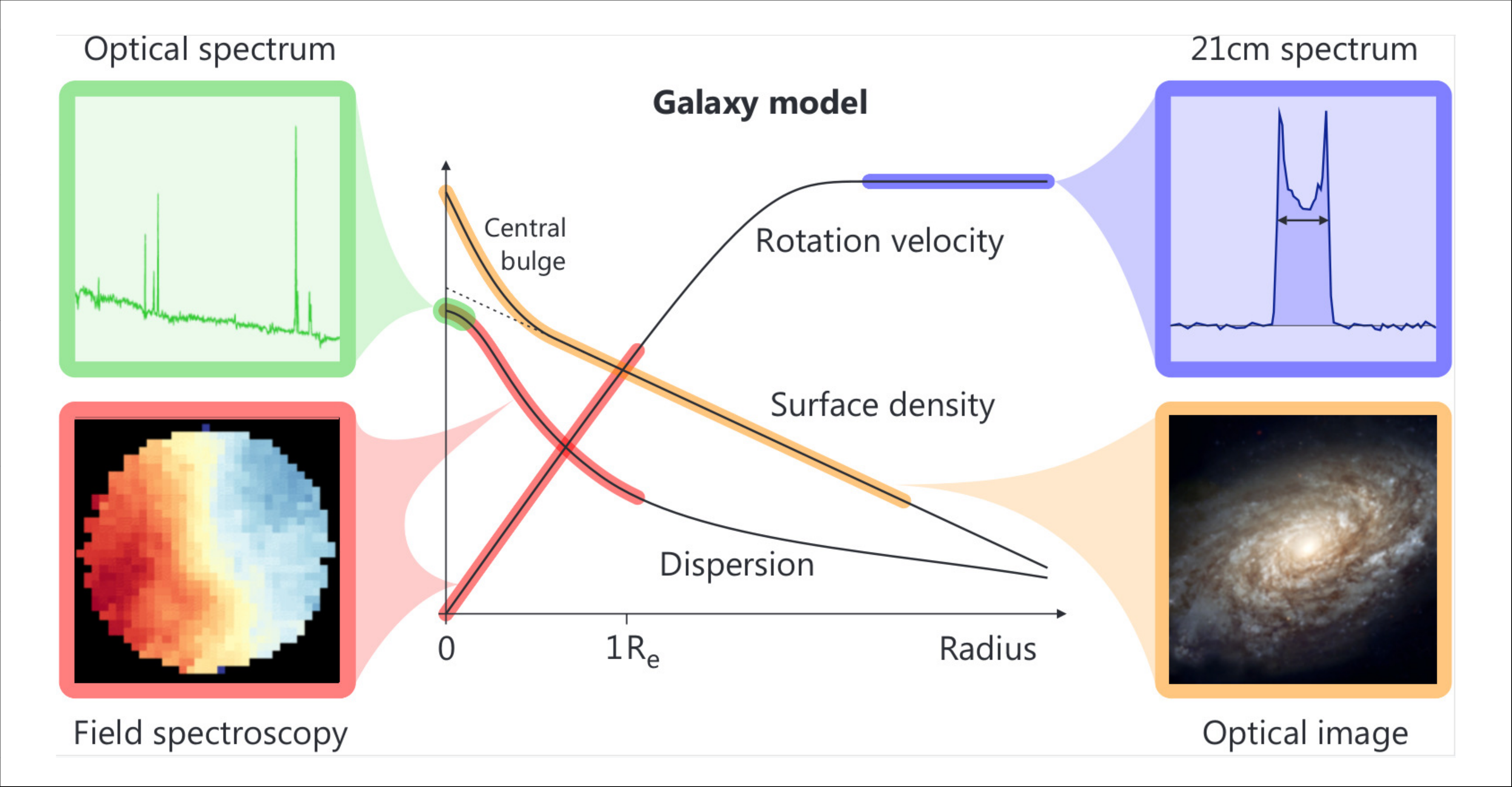}
\caption{An illustration of how various multi-wavelength data can constrain disk galaxy dynamics over different regions of a typical massive spiral galaxy with an extended disk, concentrated bulge and flat rotation curve.
\label{fig:dynamics}}
\end{figure*}

Integral-field spectroscopy (IFS) permits the inference of spatially resolved rotation and dispersion profiles by taking multiple spectra across each galaxy. Already-completed single-target surveys include SAURON \citep{deZBurEms02}, ATLAS3D \citep{CapEmsKra11}, and CALIFA \citep{SanKenGil12,SanGarZib16} - with nearly 1,000 galaxies between them - whereas ongoing multiplexed surveys like SAMI \citep{CroLawBla12,BryOweRob15} and MaNGa \citep{BunBerLaw15} have each observed $\sim$2,000 galaxies and expected to finish with 3,600/$\sim$10,000, respectively. These datasets have enormous potential to constrain fundamental galaxy properties, as illustrated in \figref{dynamics} - particularly for multi-component galaxies and when combined with multi-wavelength data like deep imaging and \HI spectra.

There are two major challenges in interpreting IFS kinematic maps. First, extracting information on galaxy kinematics requires careful modelling to account for observational and instrumental effects, particularly ``beam-smearing'' - the tendency for a point-spread function (PSF) to blur ordered rotation across a galaxy, artificially increasing the velocity dispersion. Creating spectral datacubes by stacking dithered observations has an adverse impact on image resolution, particularly in the presence of differential atmospheric refraction \citep{ShaAllFog15,LawYanBer15} - an issue which can and should be resolved by forward modelling rather than in the data reduction process. Secondly, IFS maps may not have sufficient spatial coverage or signal-to-noise to reach the peak of a typical galactic rotation curve, whereas even unresolved 21cm \HI spectra can, since \HI disks are typically more extended than stellar disks \citep[e.g.][]{WalBrideB08,WanKorSer16}.

Our new modelling method is designed to resolve the issues outlined above. We create dynamical models from fully self-consistent phase space DFs, then generate synthetic observations of the kinematic moments to compare with observed data. Using kinematic moment maps allows for less ambiguous detections of dispersion-supported bulges. Synthetic observations reproduce biases from beam smearing by the PSF/line-spread function (LSF) and pixel discretization, allowing for simultaneous fitting of independent datasets. Finally, since the models are based on theoretically-motivated analytic density profiles, they predict reasonable extrapolations beyond the limits of the observed data - vital for estimating the angular momentum in extended disks \citep{RomFal12,ObrGla14}.

Existing galaxy dynamical modelling methods include \citep{Sch79} modelling \citep[e.g.][]{CapBacBur06}, Jeans' modelling \citep[as reviewed by][]{CouCapdeJ14}, made-to-measure \citep{SyeTre96} and action-based modelling \citep[e.g.][]{BinMcM11}. However, most such methods are not specifically designed to perform bulge-disk decomposition \citep[but see][]{VasAth15} and many do not \emph{necessarily} produce self-consistent DFs \citep[e.g.][who model the Milky Way's disk DF including a halo potential but no halo DF]{TriBovRix16}. \citet{PorGerWeg17} fit Milky Way data using a near-equilibrium M2M model with a disk, halo, bulge and bar, but at a significant computational cost of 190 CPU-hours for 25 iterations. Our method solves both problems, generating synthetic observations of near-equilibrium bulge/disk/halo models efficiently enough to fit data from large surveys like SAMI.

In \secref{data}, we describe the data sources for the sample galaxy used in this pilot study. In \secref{methods}, we describe each step of the method in greater detail. In \secref{results}, we show more detailed results and comparisons to 2D bulge-disk decomposition, summarizing conclusions and outlining future directions in \secref{conclusions}. Three appendices detail systematic tests of model integration accuracy (\appref{integration}), stability (\appref{stability}) and fit robustness (\appref{fitting}). Two further appendices discuss degeneracy/biases when models fit poorly (\appref{stats}) and when fitting inclined thick disks (\appref{thickdisk}). Lastly, \appref{galactics} details the \galactics method used to build galaxy models. Future papers will provide fits to a larger sample of SAMI galaxies.

\section{Data}
\label{sec:data}

We choose a well-resolved, massive SAMI spiral galaxy (G79635), with $M_{\star} \approx 10^{10.4} \msole$ \citep{TayHopBal11} from broadband spectral energy distribution fits. Stellar kinematics are derived using single-Gaussian, two-moment pPXF \citep{CapEms04} fits to data from the blue and red arms combined, degrading the red arm (FWHM=1.696\AA, covering the redder half of SDSS r band) to match the blue arm's spectral resolution (2.717\AA, covering the SDSS $g$ band); see \citet{vdSBlaFog17} and \citet{FogScoOwe15} for details. We create ``$SAMIgr$'' flux maps by collapsing the spectral cube, masking emission and sky lines as in \citet{vdSBlaFog17}. Flux uncertainties include approximate covariances \citep{ShaAllFog15} added in quadrature to the shot/read noise, along with a flat systematic uncertainty corresponding to 10\% (2.1\%) of the faintest (peak) surface brightness. The dispersion maps exclude outliers from the best-fit radial profile. The PSF is a \citet{Mof69} ellipse with 1.83'' FWHM, derived via a \profit \citep{RobTarTob17} fit to the reference star's flux map (obtained from its spectral cube exactly as for the galaxy).

$g$- and $r$-band images are from the VST-KiDS survey \citep{deJVerKui13,deJVerBox15}, which covers GAMA \citep{DriHilKel11} and SAMI survey regions. Uncertainties are estimated from the effective gain and local sky brightness. PSFs are Moffat ellipses with 1.16'' ($g$) and 0.54'' ($r$) FWHM, derived from a simultaneous \profit fit to 39 nearby point sources. G79635 also has an \HI spectrum from the ALFALFA \citep{HayGioMar11} $\rm{\alpha}$.70 data release\footnote{\url{http://egg.astro.cornell.edu/alfalfa/data/}}.

\section{Methods}
\label{sec:methods}

The method solves a non-linear optimization problem using a parametric galaxy model, constrained by 2D kinematic moment maps or derived quantities thereof. First, a model phase space DF must be generated (\subsecref{methods_models}); second, this DF must be integrated efficiently and accurately (\subsecref{methods_dfs}); third, the integrated DF must be projected into a datacube (position-position-velocity) and then into kinematic maps (\subsecref{methods_synthetic}). Finally, the optimization and sampling procedure is described in \subsecref{methods_fits}. Our implementation - dubbed \magrite\xspace- is based on C/C++ libraries with an \R \citep{R16} interface for fitting.

\subsection{Galaxy Models}\
\label{subsec:methods_models}

The models are generated using an updated version 3.0 of the \galactics \citep{KuiDub95,WidPymDub08} galaxy initial conditions code, to be detailed in a future paper (Dubinski et al., in prep). \galactics has previously been used to model surface brightness profiles and rotation curves of local group galaxies \citep{WidDub05,WidPymDub08} and NGC6503 \citep{PugWidCou10}, but not 2D images/kinematic maps. The core functions of the updated code are as described in \citet{WidPymDub08}. Key differences include the adoption of a logarithmic grid (previously linear), and the use of GNU Scientific Library \citep{GSL09} splines to create smooth differentiable functions for tabulated DFs and multipole expansion coefficients for the potential, both of which allow for more accurate function and derivative/integral evaluations using fewer grid elements than in earlier versions.

GalactICS generates equilibrium DFs for galaxies with three components:

\begin{itemize}
\item An exponential stellar disk with mass $M_{d,in}$, scale radius $R_{d}$ and scale height $z_{d}$, where $\rho \propto \exp{(-R/R_{d})}\mathrm{sech}^2(z/z_{d})$;
\item A (deprojected) \sersic profile stellar bulge with scale velocity $v_{b}$ and effective radius $R_{b}$, where $\rho(r) \propto (r/R_{b})^{-p}\exp{(-b_{n}(r/R_{e})^{1/n_{s}})}$, $p=1-0.6097/n_{s}+0.05563/n_{s}^2$ \citep{PruSim97}, and $b_{n}$ scales such that $R_{b}$ is the projected half-light radius \citep{GraDri05}; and:
\item A generalized \citet[][hereafter NFW]{NavFreWhi97} dark matter halo with scale velocity $v_{h}$, scale radius $r_{h}$, where $\rho \propto [(r/r_{h})^{\alpha}(1+r/r_{h})^{(\beta-\alpha)}]^{-1}$, and $\alpha=1,\beta=3$ for a ``pure'' NFW profile.
\end{itemize}


The minimal set of six free parameters includes a size and mass/scale velocity per component: $R_{b}$ and $v_{b}$ (bulge); $R_{d}$ and $M_{d,in}$ (disk); $r_{h}$ and $v_{h}$ (halo). Four parameters control density profiles: $n_{s}$ (bulge), $z_{d}$ (disk scale height), and $\alpha$, $\beta$ (halo); we fix $\beta=3$ but leave the others free. We fit the disk radial \emph{central} (cylindrical) radial velocity dispersion $\sigma_{R0}$, the square of which then declines exponentially with $R_{d}$. Finally, we fix the streaming fractions $f_{s,b}$=$f_{s,h}=0.5$ (fractions of particles with positive z-axis angular momentum $L_{z}$), giving non-rotating bulges and halos.

Any component $c$ can be truncated at a radius $r_{t,c}$ with scale length $dr_{t,c}$, such that $\rho_{trunc}(r)=\rho_{nominal}(r)[1+\exp{((r-r_{t,c})/dr_{t,c})}]^{-1}$. Truncation is only strictly necessary for the halo because the NFW profile has a divergent total mass. Nonetheless, we fit the disk $r_{t,d}$ and $dr_{t,d}$ (see \secref{results} for the implications of this choice), but fix the bulge $r_{t,b}=10R_{b}$ ($dr_{t,b}=R_{b}$) and halo $r_{t,h}=50r_{s,h}$ ($dr_{t,h}=7.5r_{s,h}$). This adds an additional two free parameters to the previous nine.

\galactics derives a DF for each spherical component using Eddington's formula \citep[e.g.,][]{BinTre08} and iteratively computes corrections to an analytic DF describing the disk. \galactics then finds a potential/density pair for each component which is consistent with these DFs. The final radial density profile of each component only differs slightly from its original parametrization (see \appref{galactics}). The key differences are that spherical components (bulge and halo) are flattened slightly in response to the presence of the disk potential, and that the integrated properties of components (e.g. disk mass $M_{d}$) can deviate slightly from their input values. Although options have been recently added to \galactics to further correct the disk density such that the output mass and profile match the input values more closely, we omitted this step pending further testing of these new features.

Despite these caveats, \appref{stability} shows that under normal circumstances, \galactics models begin in near-perfect equilibrium; perturbations on the order of a few percent result only for extreme parameter combinations. More importantly, models with large \citet{Too64} $Q$ parameters are stable against secular evolution. While \galactics is not restricted to generating models with large $Q$ - this depends mostly on the values of $z_{d}$ and $\sigma_{R0}$ - it can be guided to do so if necessary through priors on these input parameters or on the minimum $Q$ value at certain radii. \galactics also converges to a near-equilibrium DF in $\sim$30 seconds without expensive orbital integration - a major advantage over Schwarzschild/made-to-measure methods and the key requirement to permit Bayesian analyses of large samples of galaxies.

\subsection{Distribution Function Integration}
\label{subsec:methods_dfs}

The default \galactics integration scheme samples the DF using a Monte-Carlo acceptance-rejection method, which is ideal for generating unbiased, equal-mass N-body particle initial conditions. However, the rejection step is inefficient unless a suitable (i.e. strictly larger than the target distribution but only by a small margin) approximate sampling distribution is known. Unbiased sampling is not optimal for accurate integration over a uniform grid, because the fractional error is not constant but scales with density, so low-density (outer) regions have larger relative errors than the (possibly excessively) accurately-sampled inner regions. Lastly, stochastic integration can induce spurious variations in the likelihood with small changes in parameter values. Evaluating the same model with a different random sequence or even a slightly different model with an identical random seed can result in spurious differences in integrated quantities and the resulting model likelihoods, depending on the number of samples. As a result, we chose to develop a faster and less stochastic grid-based integration scheme which we will now describe in greater detail. For more detailed comparisons between these two integration schemes, see \appref{integration}.

We integrate the DF in its native cylindrical coordinate system and then project it rather than integrating over projected coordinates, as is usually done for 2D surface brightness profiles. For the remainder of this section, we will use the mathematical convention where the azimuthal angle is $\theta$ rather than the physics convention ($\phi$). The disk DF is parametrized as $f(R,z,v_{R},v_{\theta},v_{z})$. It is independent of $\theta$ and symmetric over all axes except $v_{\theta}$. The DFs of spherical components (bulge and halo) are internally functions of energy and $L_{z}$, but are re-parametrized as $f(R,z,v)$ for convenience. Integrating the model over a cylindrical grid allows for some efficient optimizations, whereas integrating down the line of sight requires repeated unique transformations at each projected position. Rotating and projecting cylindrical grid elements into the sky plane does present a challenge. Doing so exactly requires computing the fraction of the volume of a 3D tilted ring (with a fixed height) projected within each spaxel. However, this can be roughly approximated by further discretizing each annular ring into sectors, and then assigning the mass within each sector to the spaxel containing its center of mass (in projection).

\begin{figure*}
\includegraphics[width=\textwidth]{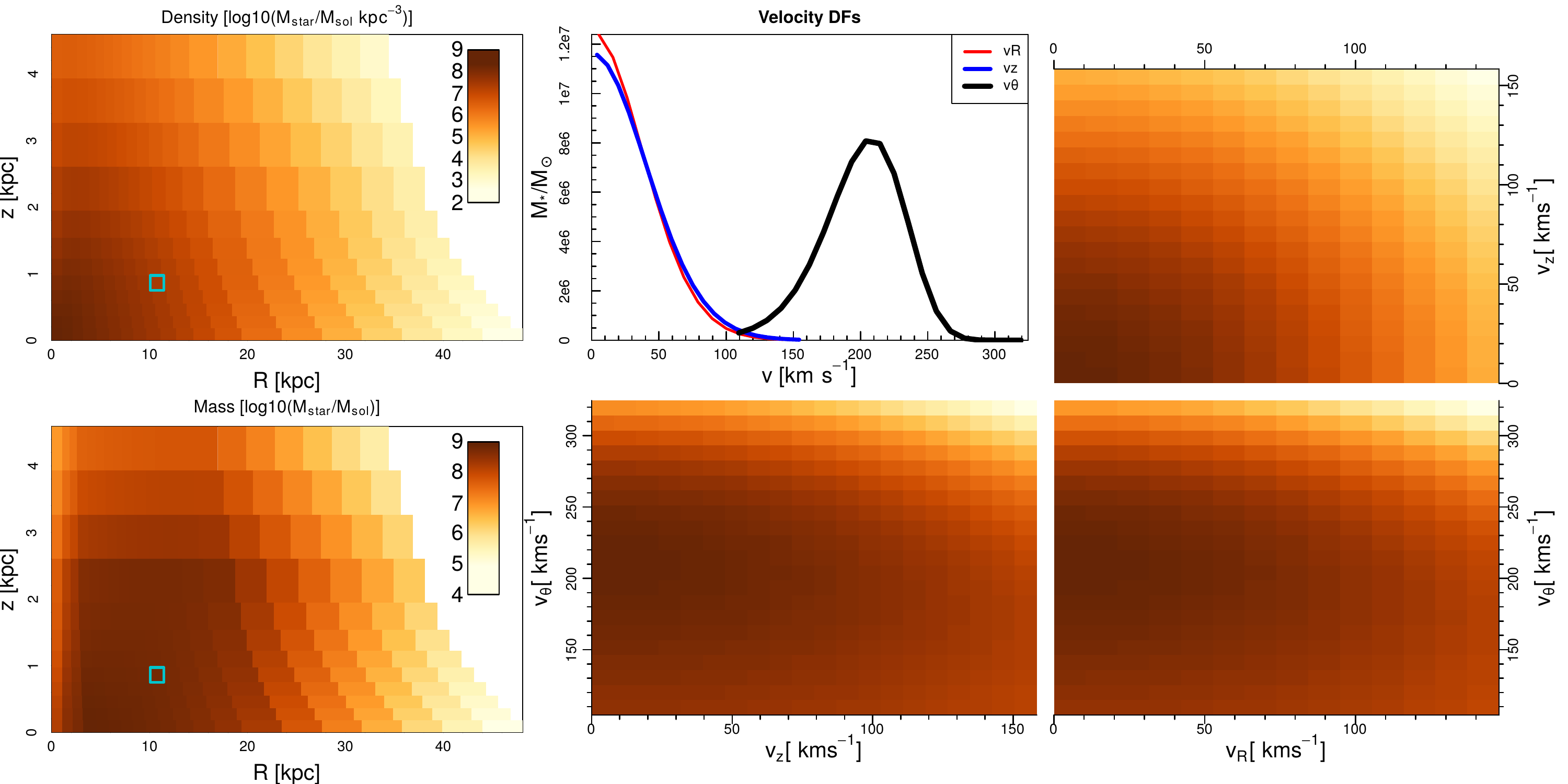}
\caption{Densities within selected bins of a low-resolution model grid (25 radial bins and $25\kmse$ velocity bins). Left panels: Density and mass within the model integration grid. The scheme produces roughly equal-mass bins over most of the disk. Top: Mass-weighted velocity distribution functions for the cyan-highlighted bin in the left panels. The remaining panels show pairwise (logarithmic) densities integrated over the third velocity axis, i.e. projections of the velocity ``ellipsoid'' (which is not perfectly ellipsoidal) at a given position in the disk.
\label{fig:integration_grid}}
\end{figure*}

The discretized disk integration grid for G79635 is shown in \figref{integration_grid} (top-left panels). The scheme is designed to create nearly equal-mass bins. The radial grid is roughly logarithmic - each bin covers a radius containing 1/$N_{R}$ of the total mass of an ideal, thin exponential disk. The inner and outer bins are oversampled to minimize gaps at large radii and improve accuracy near the galactic center: 15\% of the bins cover the inner 0.5$R_{d}$, whereas 35\% cover $R>3R_{d}$. The bins are staggered radially to spread them more evenly in projection. Vertically, the grid covers $0<z<10z_{d}$, spaced to cover equal masses until switching to linear spacing near the upper limit. For each $R$-$z$ element, the disk DF is integrated over all $v_{R},v_{z},v_{\theta}$ within $(\mean{v_{R}}=0)\pm4\sigma_{R}$, $(\mean{v_{z}}=0)\pm4\sigma_{z}$ and $(\mean{v_{\theta}}\approx v_{circ})\pm8{\sigma_{\theta}}$, discretized into equal-velocity bins. \figref{integration_grid} shows the integration grid for a single spatial bin, including major-axis 2D projections and 1D probability distribution functions (PDFs). Typically, the DF at most spatial coordinates in the disk is nearly (but not exactly) a Gaussian ellipsoid.

The bulge uses similar radial divisions, such that the inner and outer 20\% of the bins contain $0.1M_{bulge}$ and $0.05M_{bulge}$, respectively, accounting for the steep slope in the \sersic profile at small/large radii for large/small values of $n_{s}$, respectively. The radial grid is divided into quadrants and then subdivided in linearly-spaced cells along the z-axis. The bulge DF is then integrated over all $v<v_{esc}$.

\subsection{Synthetic Observation Pipeline}
\label{subsec:methods_synthetic}

To generate synthetic images and kinematic maps, we use an updated version of the synthetic observation pipeline described in \cite{TarDubYee13} and ironically named ``This Is Not A Pipeline'' (TINAP). We assume an exponentially-declining star formation history for the disk: $SFR\propto\exp{(-(t-t_{0})/\tau)}$ from $t_{0}=2$~Gyr to 12.92~Gyr (the Universe's age at G79635's $z=0.04$ assuming $H_{0}=70~\kmse \mathrm{Mpc^{-1}}$, $\Omega_{m}=0.3$, $\Omega_{\lambda}=0.7$), fitting $\tau^{-1}$ to avoid discontinuities at $\tau=0$. The bulge is modelled as a single burst with a free formation time $t_{b}$. Both bulge and disk components have free metallicities ($Z_{b}$ and $Z_{d}$, respectively). We use \citet{MarStr11} grids to compute $M_{\star}/L$ in the three bands ($g$, $r$ and $SAMIgr$), assuming no stellar population gradients within components. We spawn a minimum of 8 ``particles'' at the central $R,z$ of each grid bin with an evenly-spaced distribution from $0<\theta<\pi/4$, beginning at a random $\theta$ (the only stochastic part of the scheme) and duplicating particles in the seven remaining octants.

\begin{figure*}
\includegraphics[width=\textwidth]{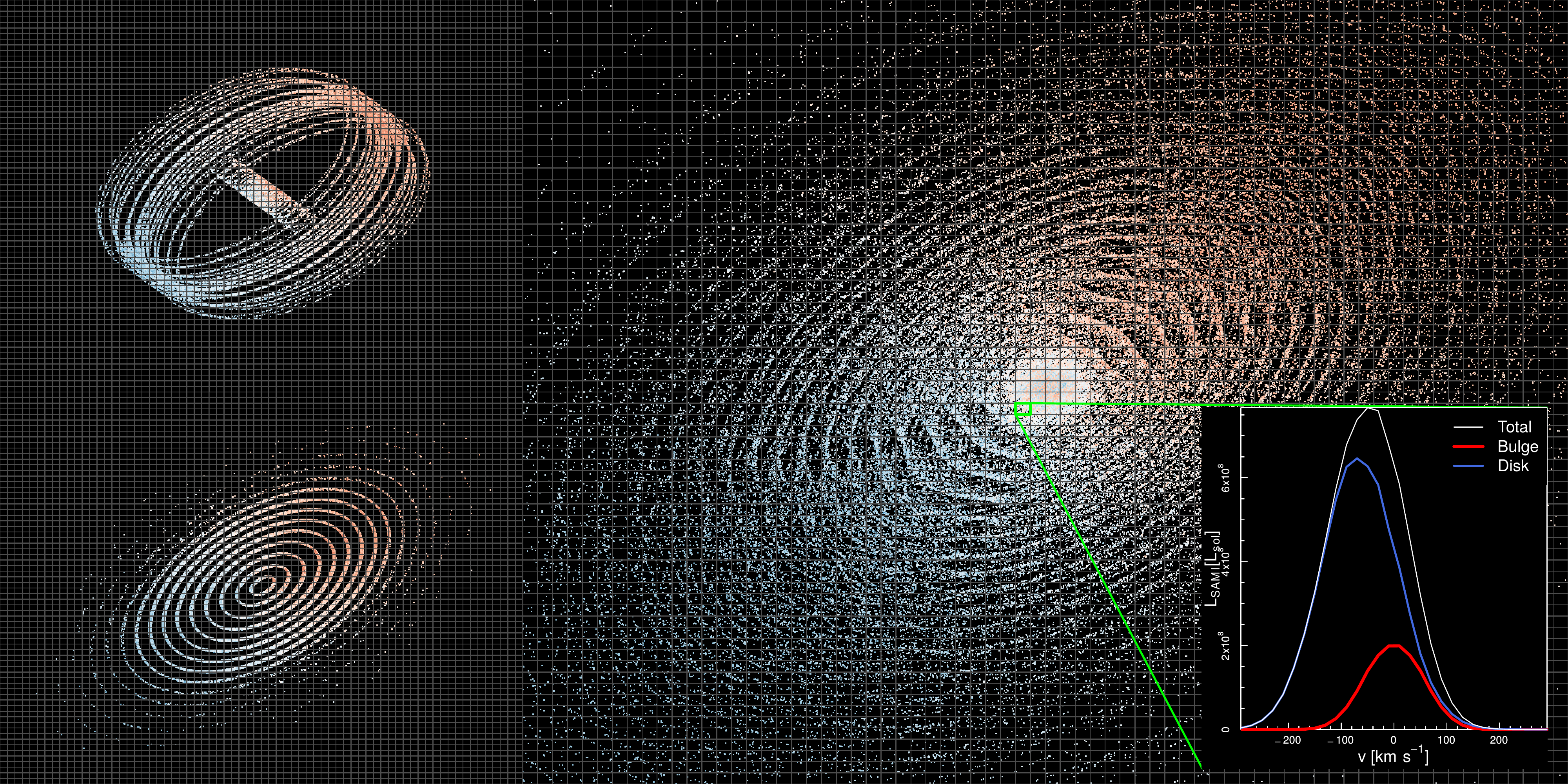}
\caption{A visualization of the model projection and map generation schemes. Left panels: Scatter plots of projected disk DF samples, color-coded by line-of-sight velocity ($v_{LOS}$). The top-left panel shows bins at similar radii but different heights above/below the midplane, while the bottom-left panel shows a single $z$ bin with ellipses at same height above/below the midplane but different radii. Right panel: Scatter plot of DF samples for the disk and bulge, color-coded by $v_{LOS}$. The inset shows $v_{LOS}$ DFs for the highlighted spaxel (green). Radially-staggered bins are distributed more evenly in projection but still create artifacts.
\label{fig:integration_projected}}
\end{figure*}

The two left panels of \figref{integration_projected} show distributions of disk particles at two fixed radii but at different heights above/below the disk midplane, colour-coded by $v_{LOS}$ (top), along with particles at different radii but fixed heights above/below the disk midplane (bottom). After binning particles spatially and in $v_{LOS}$, every spaxel produces its own $v_{LOS}$ PDF (right panel inset, \figref{integration_projected}). These PDFs are 2D integrals of line-of-sight projections of the 3D velocity ellipsoids, and so kinematic moments are sensitive to the disk's vertical structure and anisotropy. It is worth emphasizing that \figref{integration_grid} shows a very coarse integration grid with just 25 radial bins, whereas for G79635 we use 100 bins, eliminating most discreteness effects. However, the X-shaped pattern of gaps remains even for very fine grids. This is essentially a Moir\'{e} pattern generated by overlaying an elliptical grid onto a rectangular one. The effect is minimized but not eliminated by staggering radial bins. In practice, the patterns are small enough to be virtually invisible after PSF convolution and could be avoided entirely with more sophisticated schemes for gridding the model DF, which are under development.

For stellar kinematics, $v_{LOS}$ cubes are convolved with the PSF and spectral line spread function (LSF), both of which are oversampled threefold. Finally, we measure the kinematic moments in each spaxel, subtracting the LSF dispersion in quadrature for the second moment. Gaussian fits to G79635's $v_{LOS}$ PDFs are indistinguishable from direct measurements of $v_{LOS},\sigma$, so we use the latter. As discussed in \appref{fitting}, this choice may not be suitable for galaxies with more massive and extended bulges, so we are investigating alternatives for future applications.

Generating synthetic kinematic maps requires nine additional free parameters - the disk inclination and position angle, offsets for $x,y,v_{los}$, and ages and metallicities for the bulge and disk - bringing the total to twenty one free parameters.

\subsection{Model Fitting}
\label{subsec:methods_fits}

Wherever possible, initial parameter estimates and prior means are obtained from 2D \profit fits. All priors are assumed to be normal and broad ($\sigma\approx1$ dex). We first use a robust maximum-likelihood genetic algorithm, Covariance Matrix Adaptation - Evolutionary Strategy \citep[\cmaes;][]{cmaes}. We then perform Bayesian MCMC using the Componentwise Hit-And-Run Monte-Carlo (CHARM) sampler of the \laplacesdemon\xspace\R package\footnote{\url{https://cran.r-project.org/web/packages/LaplacesDemon/}}. The likelihood function is the sum of the log-likelihoods from each map, assuming either a chi-square distribution for image residuals, or a sum of normally-distributed residuals for kinematic maps with less well-defined errors. Residuals are defined as $\chi_{i} = (data_{i}-model_{i})/error_{i}$ and $\chi^2 = \Sigma_i ((data_{i}-model_{i})/error_{i})^2$. Note that although we quote reduced $\chi^2$ ($\chi^2_{red}$) values, they are not used in the optimization procedure, which instead derives likelihoods from the chosen statistics' PDF directly (i.e. by calling the ``dnorm'' and ``dchisq'' functions in \R).

Our \cmaes code is based on the \R cmaes package\footnote{\url{https://cran.r-project.org/web/packages/cmaes/}}. We have modified both \cmaes and \laplacesdemon, implementing runtime limits for supercomputer queues; these versions are available on github\footnote{\url{https://github.com/taranu/LaplacesDemon}}$^{,}$\footnote{\url{https://github.com/taranu/cmaeshpc}}.

\section{Results}
\label{sec:results}

\begin{figure*}
\includegraphics[width=\textwidth]{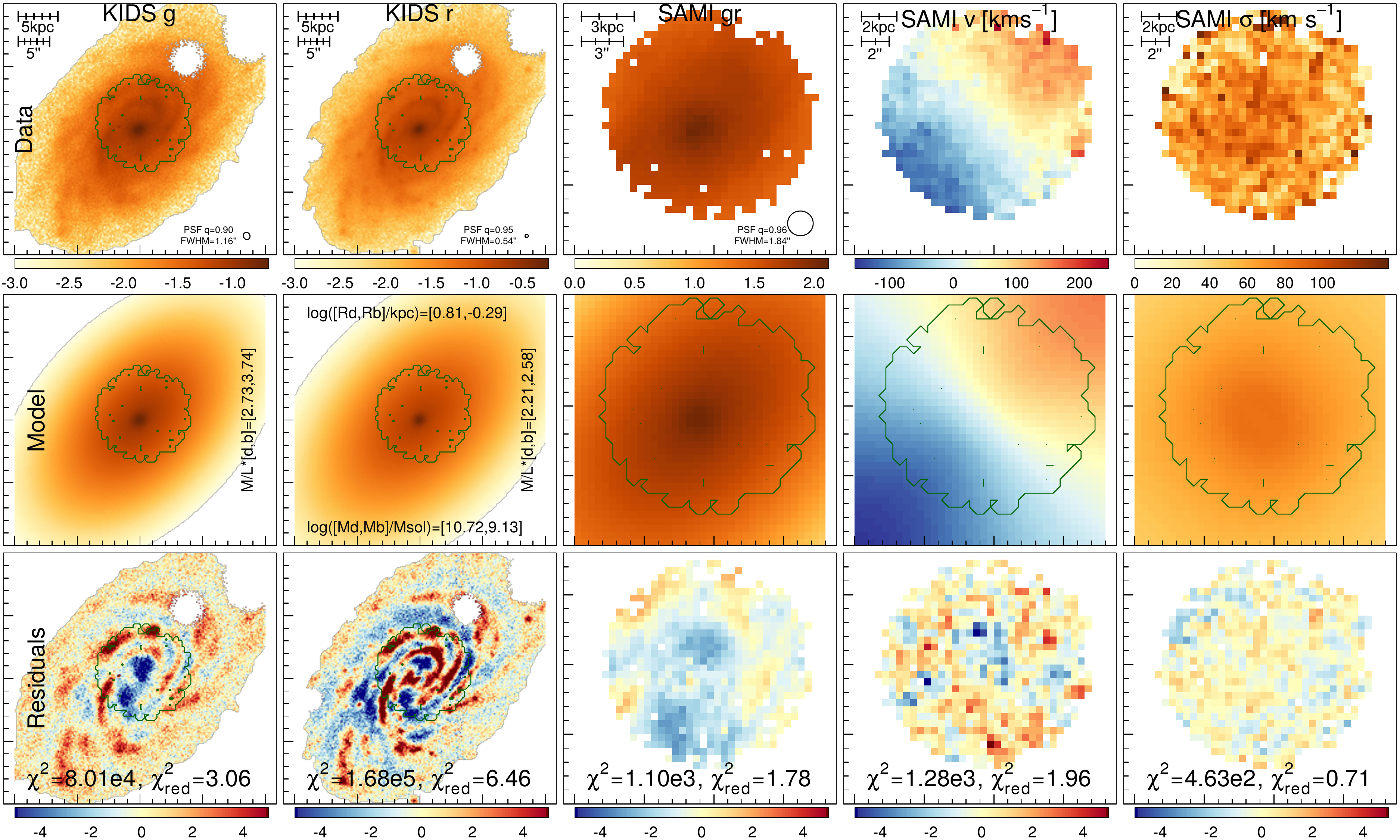}
\caption{Best-fit G79635 model using SAMI moment 0-2 maps and KiDS $g+r$ images, along with residuals relative to pixel/spaxel uncertainties. KiDS images are 200x200 ($2''$ ticks) while SAMI are 36x36 ($1''$ ticks).
\label{fig:bestfit_samikids_2D}}
\end{figure*}

\begin{figure*}
\includegraphics[width=\textwidth]{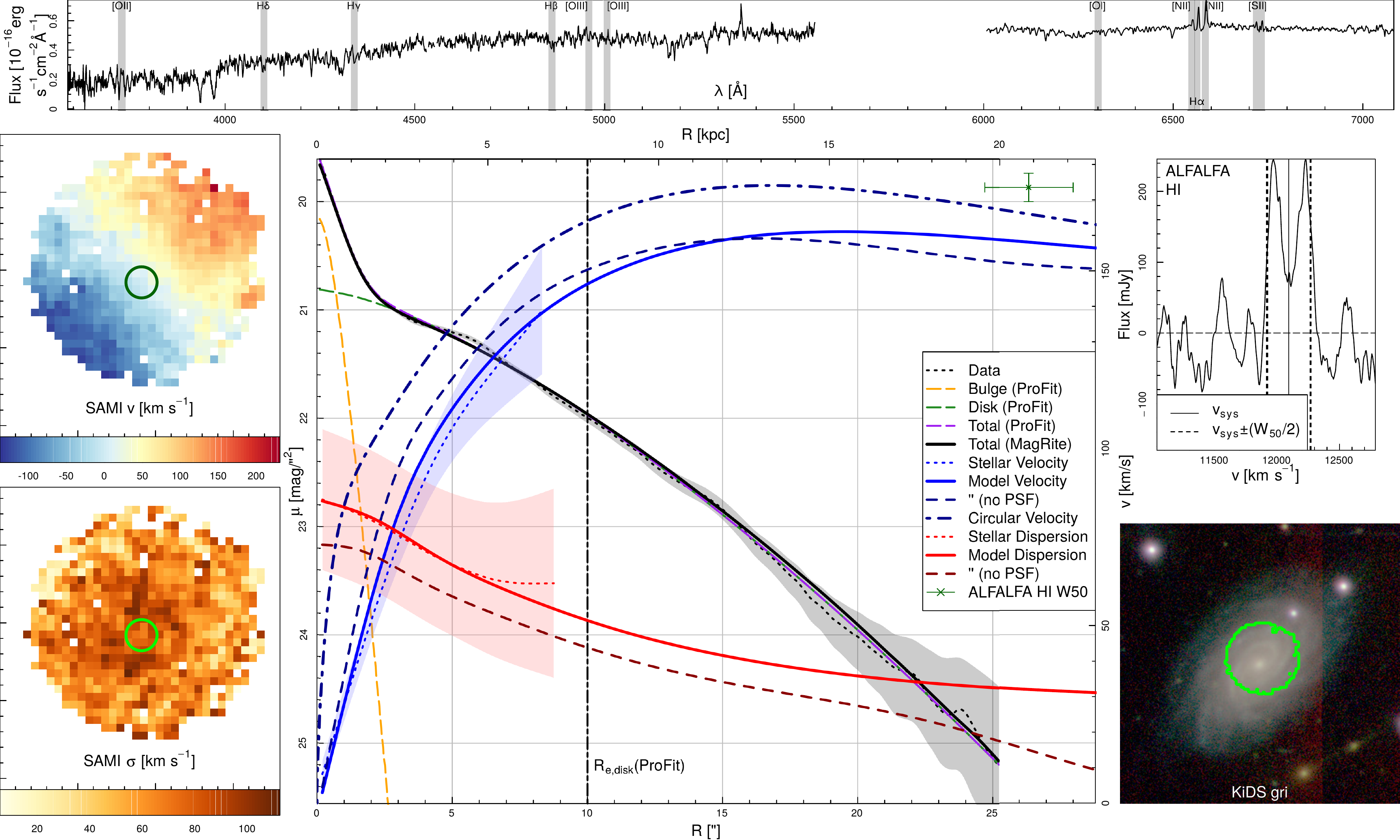}
\caption{Data and 1D profiles for G79635. Clockwise from bottom-left: the SAMI dispersion and velocity maps, with a 2" diameter aperture (green); the SAMI spectrum within this aperture, with emission lines excluded from the fit shaded in gray; the ALFALFA \HI spectrum; RGB image using KiDS $g$- and $r$-band images, overlaid with SAMI coverage (green). Center: azimuthally-averaged, spline-fit 1D profiles (dotted lines) for the first three kinematic moments (KiDS $r$-band surface brightness - gray/black; velocity - blue; velocity dispersion - red), along with 1-$\sigma$ uncertainties (shaded). Also shown are best-fit \magrite (thick solid lines) and \profit 2D double \sersic r-band fits (dotted lines), split between bulge (orange), disk (green), and total (purple). Note that velocities are as observed and not inclination-corrected, i.e. $v \approx v_{circ}\sin{(i)}$, where $v_{circ}$ is the circular velocity and $i$ is the disk inclination. The best-fit \magrite model $v_{circ}\sin{(i)}$ (dot-dashed dark blue line) lies well above the observed rotation curve, illustrating the combined effects of beam smearing, asymmetric drift and a thick disk.
\label{fig:bestfit_samikids_1D}}
\end{figure*}

The best fit for G79635 using SAMI and KiDS $g+r$ is shown in \figref{bestfit_samikids_2D}. The $\chi^2_{red}$ for all of the flux maps is significantly above unity. However, the largest $r$-band residuals clearly trace non-axisymmetric features like spiral arms and inter-arm gaps, and the similarity in residuals across independent data sets is encouraging, given the systematics introduced by SAMI's cubing procedure and single-star flux calibration. The $r$-band fit is worst simply due to its higher signal-to-noise (better seeing and longer exposures than $g$).

\figref{bestfit_samikids_1D} shows 1D profiles azimuthally averaged over the best-fit \profit disk ellipse, compared to a \profit 2D double \sersic r-band fit with a free bulge position angle. The dispersion map/profile is overfit and the best-fit rotation curve appears to rise slightly too steeply, as can also be seen in \figref{bestfit_samikids_2D} (where the velocity map residuals show spatial coherence). Encouragingly, the predicted rotation curve at a fiducial radius of (3--3.4)$R_{d}$ \citep{CatHayGio07} is within 10\% of the independent ALFALFA \HI $W_{50}=(347 \pm 8)\kmse$ measurement, even though the \HI data was not used in the fit and the SAMI data does not appear to reach the peak of the rotation curve. The lower stellar velocity could be due to asymmetric drift, as it is not unusual for stellar disks with radial dispersion support to have $\sim10\%$ lower rotation speeds than gaseous disks \citep{CiaDurLay04,MarVerWes13,BroPapChr17}. The peak stellar velocity is also consistent with the independent $V_{max}\sin{i}=165\kmse$ circular speed derived by \citet{CecFogRic16}.

The fact that the observed mean stellar velocity lies well below the circular speed curve is due to a combination of factors. First, the mean velocity within $R_{e,disk}$ is decreased due to beam smearing (compare the solid blue and dashed blue lines in \figref{bestfit_samikids_1D}). This effect is modest beyond the peak of the rotation curve (compare the solid and dashed rotation curves), although it continues to boost the observed velocity dispersion by about 10$\kmse$. Note that estimates of the mean velocity and dispersion are unreliable beyond about 15kpc, where the dispersion drops well below the 60$\kmse$ velocity grid resolution. Also, there is some subjectivity in how 1D apertures are defined. We measure velocities and velocity dispersions using data from spaxels within 5 and 10 degrees of the major axis, respectively. The model projected velocity without beam smearing (dashed blue curve in \figref{bestfit_samikids_1D}) is measured within the same apertures as the PSF-convolved version (solid line) and does not take into account the fact that PSF convolution modifies isophotes as well; however, since the 1D kinematics are measured close to the major axis, this effect is minor.

In the inner few kpc, the mean velocity is suppressed both because of the presence of a non-rotating bulge and because the disk has a finite thickness, so that a large fraction of disk stars are a significant distance away from the disk midplane. Beyond this inner region, asymmetric drift and the non-zero radial and vertical velocity dispersion of the disk continue to lower the mean velocity. Observations suggest that it is not unusual for stellar disks with radial dispersion support to have $\sim10\%$ lower rotation speeds than gaseous disks \citep{CiaDurLay04,MarVerWes13,BroPapChr17}. 

Despite also fitting the $g$-band image and SAMI kinematics, the \magrite best-fit model is a better fit to the KiDS $r$-band image than a single-band, exponential disk \profit fit; \profit only fits slightly better with a free $n_{s}$ disk.

\begin{table}[htbp]
\caption{Best-fit G79635 \magrite Parameters}
Fitted Parameters\\
\\
\begin{tabular}{ccccccc}
\hline
Name & unit & $^{1}$log & mean & $\sigma_{obs.}$ & $\sigma_{mock}$ & $^{2}\Delta$\\
\hline 
$M_{d,in}$ & $\mathrm{^{3}M_{sim}}$ & Y  & 1.604 & 8.76e-5 & 1.33e-3 \\
$R_{d}$ & kpc & Y  & 8.141e-1 & 2.62e-5 & 9.85e-4 \\
$z_{d}$ & kpc & Y  & 2.224e-1 & 1.83e-4 & 8.08e-3 \\
$r_{t,d}$ & kpc & Y  & 1.146 & 2.07e-4 & 1.67e-3 \\
$dr_{t,d}$ & kpc & Y  & 7.065e-1 & 3.56e-4 & 5.26e-3 \\
$\sigma_{R0}$ & $\mathrm{^{4}v_{sim}}$ & Y  & -1.281e-1 & 2.93e-4 & 2.44e-2 \\
\hline 
$v_{b}$ & $\mathrm{\sqrt{v_{sim}}}$ & N  & 1.020e-2 & 1.80e-4 & 1.92e-3 \\
$R_{b}$ & kpc & N  & -2.899e-1 & 1.01e-4 & 7.35e-3 & -3.62e-2\\
$n_{s}$ & N/A & Y  & -9.970e-2 & 9.79e-4 & 2.45e-2 & 1.82e-1\\
\hline 
$v_{h}$ & $\mathrm{\sqrt{v_{sim}}}$ & Y  & 2.693e-1 & 6.41e-5 & 1.82e-3 \\
$r_{h}$ & kpc & Y  & 8.051e-1 & 3.17e-4 & 1.47e-2 \\
$\alpha$ & N/A & N  & 9.845e-1 & 1.57e-4 & 2.31e-2 \\
\hline 
$\tau^{-1}$ & Gyr$^{-1}$ & N  & 4.851e-1 & 1.36e-3 & 6.62e-3 \\
$t_{b}$ & Gyr & Y  & 8.450e-1 & 1.19e-3 & 8.55e-3 \\
$Z_{d}$ & $\mathrm{Z/Z_{\odot}}$ & Y  & -5.170e-1 & 1.75e-3 & 7.19e-3 \\
$Z_{b}$ & $\mathrm{Z/Z_{\odot}}$ & N  & 3.000e-1 & 8.17e-5 & 8.83e-3 \\
\hline
P.A. & rad. & N  & 8.059e-1 & 2.72e-4 & 2.08e-3 & -3.26e-3\\
$\sin{(i)}$ & rad. & N  & 8.160e-1 & 1.55e-4 & 1.65e-3 & 2.09e-2\\
$X_{off}$ & kpc & N  & 3.732e-2 & 4.32e-4 & 5.81e-3 & 5.11e-2\\
$Y_{off}$ & kpc & N  & 1.513e-2 & 9.68e-4 & 5.89e-3 & 5.91e-2	\\
$V_{z,off}$ & \kms & Y  & -1.547e-1 & 3.47e-1 & 3.55e-1 \\
\end{tabular}
\newline \\
Derived Parameters \\
\\
\begin{tabular}{ccccccc}
\hline
Name & unit & $^{1}$log & mean & $\sigma_{obs.}$ & $\sigma_{mock}$ & $^{2}\Delta$\\
\hline 
$^{5}M_{d}$ & \msol & N  & 10.72 & 1.43e-4 & 1.38e-3 & \\
$M_{b}$ & \msol & N  & 9.126 & 6.93e-4 & 1.07e-2 & \\
$(M/L_{g})_{d}$ & $\mathrm{(M/L_{g})_{\sol}}$ & N  & 2.726 & 1.47e-3 & 1.18e-2 & \\
$(M/L_{r})_{d}$ & $\mathrm{(M/L_{r})_{\sol}}$ & Y  & 2.213 & 8.60e-4 & 7.42e-3 & \\
$(M/L_{g})_{b}$ & $\mathrm{(M/L_{g})_{\sol}}$ & N  & 3.745 & 1.18e-2 & 8.79e-2 & \\
$(M/L_{r})_{b}$ & $\mathrm{(M/L_{r})_{\sol}}$ & N  & 2.581 & 7.44e-3 & 5.42e-2 & \\
$L_{d}$ & $\mathrm{L_{\sol,r}}$ & Y  & 10.38 & 1.54e-4 & 8.76e-4 & 4.46e-3\\
$L_{b}$ & $\mathrm{L_{\sol,r}}$ & Y  & 8.715 & 9.91e-4 & 7.58e-3 & 2.99e-2\\
$^{4}R_{e,d}$ & kpc & Y & 0.9365 & 8.82e-5 & 1.10e-3 & 7.10e-3\\
$M_{h}$ & \msol & Y & 11.81 & 3.03e-4 & 3.33e-2 & \\
\end{tabular}
\tablecomments{Fitted parameters are listed as fit internally by \magrite, whereas derived parameters are measured during or after model generation. \\
$^{1}$ Values listed as $\log10{(\mathrm{value}~\mathrm{unit}^{-1})}$\\
$^{2}\Delta$: \magrite value less \profit value (where applicable), where the \profit model has a thin \sersic disk sharing its position angle with the bulge.\\
$^{3}$ $\mathrm{M_{sim}}$=$2.3245$e$9~M_{\odot}$\\
$^{4}$ $\mathrm{v_{sim}}$=$100~\kmse$\\
$^{5}$ Directly measured model half-light radius, accounting for truncation
}
\label{tab:fitvalues}
\end{table}

\begin{figure*}
\includegraphics[width=\textwidth]{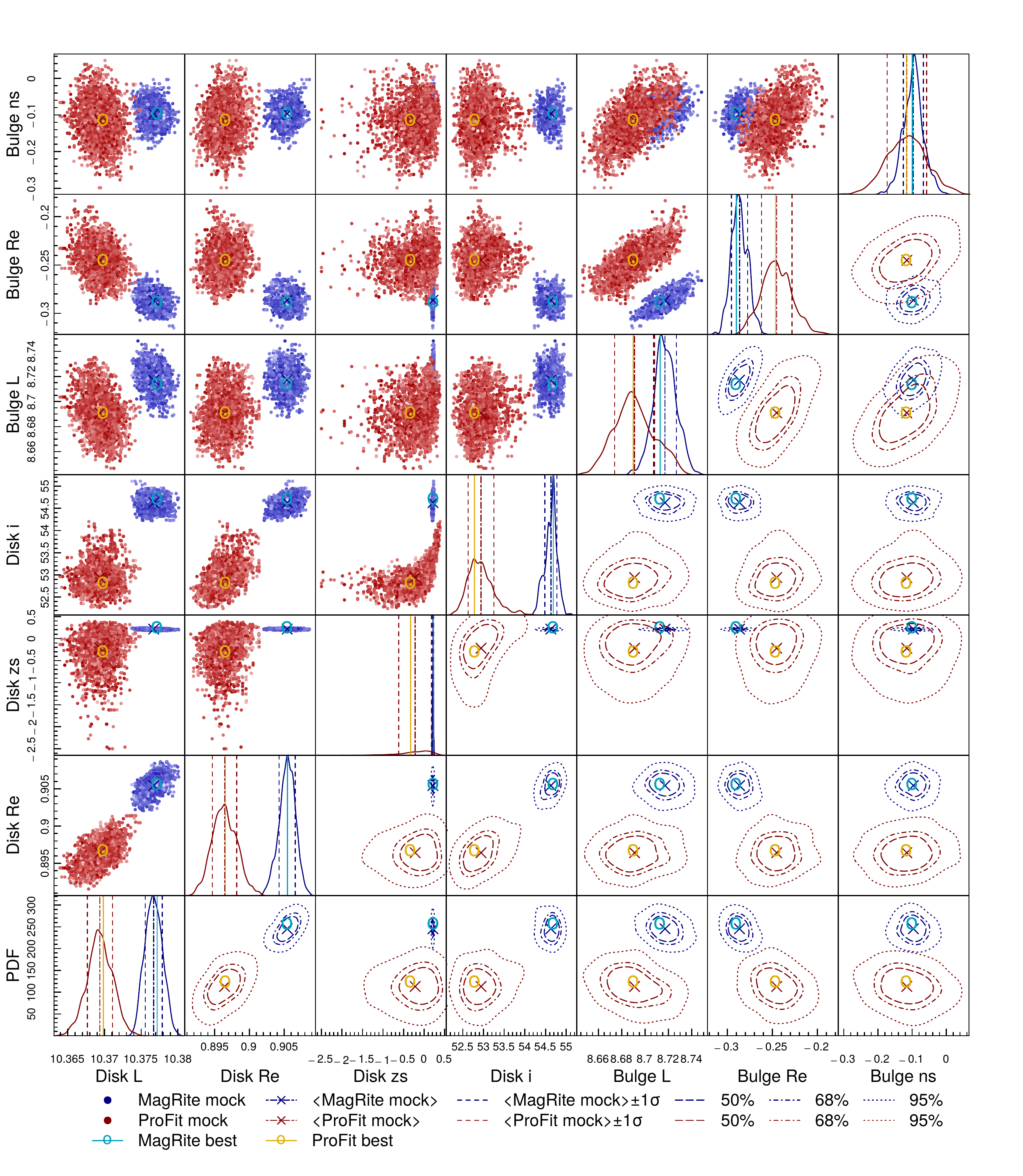}
\caption{Triangle plot showing joint posterior parameter distributions ($L\equiv\log10{(L_{r}/\lsole)}$, $Re\equiv\log10{(R_{e}/\mathrm{kpc})}$, $zs\equiv\log10{(z_{d}/kpc)}$, $n\equiv\log10{(n_{s})}$, and $i$ is the inclination in degrees) for \profit (blue) and \magrite (red), where the \profit disk is a \sersic profile and shares its position angle with the bulge. The upper-left half shows scatter plots of accepted samples, while the bottom-right half shows 1D and smoothed 2D probability contours. Accepted samples are colored by probability on an arbitrary scale, such that more probable points have darker and more saturated colors. Plots along the diagonals show PDFs of accepted samples for the variable on the x-axis; to avoid crowding, the y-axis ticks and labels are omitted for the three interior histograms. All posterior distributions are for fits to mock data; the best-fit parameters used to generate the mock data are plotted as circles. Note that because the \profit thick disk fits have an unrealistically large disk scale height, the \profit mock uses the best-fit thin disk fit parameters with $z_{d}=0.1R_{e,d}/1.67835$ (equivalent to $0.1R_{d}$ for an exponential disk); this illustrates the limited constraints on disk thickness from photometry alone.
\label{fig:mcmc_samikids}}
\end{figure*}

\tabref{fitvalues} lists best-fit values and uncertainties for \magrite model parameters and several key derived quantities. Posterior distributions for selected common parameters of the \magrite and \profit fits are shown in \figref{mcmc_samikids}. We find that direct fits to the data yield unreasonably narrow PDFs, listed as $\sigma_{obs.}$ in \tabref{fitvalues}. To test whether \magrite is the cause of this effect, we generate and fit noisy mock maps of the best-fit model (see \appref{fitting} for a full description of the procedure). We find no evidence for significant bias in the best-fit parameter values. This form of ``noise bias'' can be significant in low signal-to-noise image, as is the case in weak lensing studies \citep[e.g][]{BerJar02,RefKacAma12}. However, the parameter PDFs for fits to the mock maps are significantly broader (\tabref{fitvalues}) than when fitting the actual data - in some cases by over two orders of magnitude - and \profit exhibits similar behaviour.

As \figref{bestfit_samikids_2D} shows, an axisymmetric disk is not a good fit to the flux maps and cannot reproduce the spiral arm structure evident in the KiDS images (especially in $r$). In general, models that fit data poorly underestimate uncertainties significantly, although the degree to which this occurs does depend on the model and fit statistic. This result is not immediately obvious and we discuss it further in \appref{stats}. Our solution of fitting mock images to obtain more realistic parameter uncertainties is necessary but likely insufficient. That is, the $\sigma_{mock}$ in \tabref{fitvalues} should be interpreted as a lower bound on the uncertainty on each parameter in the highly idealized scenario that the galaxy is perfectly described by the model. There is no obvious prescription for estimating or adjusting parameter uncertainties for models that do not fit data well.

\tabref{fitvalues} also lists $\Delta$, the difference between the \profit and \magrite best-fit values for common parameters (whether derived or fit directly). This can be considered as an estimate of systematic uncertainties from using two different (but still similar) modelling methods. In all cases, $|\Delta|$ is larger than $sigma_{mock}$ - sometimes by more than an order of magnitude. This suggests that systematic uncertainties dominate over statistical uncertainties. Unfortunately, we are unaware of any robust methods for incorporating systematic uncertainties into our likelihood functions, so the only obvious solution to this issue remains increasing the model's flexibility until it can reproduce the data.

Our testing demonstrates that \magrite will recover input parameters correctly from idealized mock data, but this does not guarantee realistic parameter values when fitting real galaxies. For example, the \magrite model has an unrealistically large disk scale height $z_{d}=1.67$kpc and a small truncation radius $r_{t,d}=14.0$kpc as compared to the scale length $R_{d}=6.52$kpc. These values seem to compensate for features in the data not otherwise described by the model. G79635's disk appears steeper than exponential, and the best-fit \profit disk $n_{s}\approx0.8$ (\figref{bestfit_samikids_1D}). The small truncation radius steepens the \magrite surface brightness profile at large radii, whereas the large scale height lowers the surface brightness along the minor axis from the galaxy center - precisely where there are two under-dense inter-arm gaps. A model with azimuthal variations and a more general \sersic or broken-exponential disk profile might prefer a thinner, non-truncated disk. Having said that, \citet{MunSheGil13} fit broken exponential profiles to Spitzer 3.6\si{\micro\metre} imaging of nearly face-on disks and found a typical break radius at $2.3\pm0.9$ inner scale lengths, so the truncation radius is not unreasonable for a Type II \citep[][truncated, as per]{Fre70} disk.

The disk mass is also considerably higher than the total stellar mass estimated by \cite{TayHopBal11} from fits to photometry alone. G79635 has a rather large estimated \HI mass of $10^{10.22}\msole$, so it is possible that our larger disk mass is compensating for the contribution of the gas disk to the rotation curve. This could also be the cause of the slight under-prediction of the rotation curve at large radii, if it is not due to asymmetric drift. In practice, a more flexible and better-fitting stellar mass model would likely allow the halo parameters to vary more to compensate for such inconsistencies.

The best-fit disk metallicity is quite low ($\log10{(Z_{d}/\zsole)}<-0.5$) for such a massive disk, whereas the bulge metallicity reaches the ceiling of the \citet{MarStr11} model grids ($\log10{(Z_{b}/\zsole)}=0.3$). By contrast, the disk is fairly old, with a short $\tau=2.06$~Gyr, while the bulge has a moderate age of 5.92~Gyr. The observed galaxy colours cannot be reproduced by such a relatively simple model; in particular, the galaxy center is redder than the model, and the outskirts are significantly bluer, both by about 0.2 in $g-r$ and with a fairly sharp transition rather than a smooth gradient. Additional model complexity (especially dust reddening and stellar population gradients) is necessary to fully reproduce galaxy colours, and full spectral modelling would be ideal. However, it is worth noting that systematic differences in the inferred stellar masses of GAMA galaxies (including the SAMI sample) can be as large as 0.2 dex depending largely on the treatment of star formation histories and dust \citep{WriRobDri17}, even neglecting possible variations in the initial mass function. There are also significant differences between stellar population models, stellar spectral libraries and isochrones which preclude making accurate estimates of stellar mass-to-light ratios even given a star formation history and it is unclear how one might estimate the magnitude of such effects for a given galaxy.

One potential general solution to limit parameter bias is to introduce stricter priors on model parameters based on external data. Disk scale height distributions can be constrained from independent observations of edge-on disks, and dust extinction can be estimated from Balmer line decrements. Ultimately, the best solution is to improve the model itself, which would permit quantification of such biases. Such improvements are planned for \magrite but are not necessary to implement the method itself. For the moment, we advise caution when interpreting uncertainties from models that do not adequately reproduce known or clearly visible features in the data. This particular galaxy is well-resolved compared to average SAMI galaxies (although not uniquely so); these issues are less pronounced when fitting lower-quality data. However, as the recent public release of Subaru Hyper Suprime-Cam data \citep{AihArmBic17} shows, high-quality, deep ground-based imaging is rapidly becoming available for large galaxy surveys - even in the southern sky \citep{KelSchBes07} - and so this issue cannot be ignored for much longer.

\section{Conclusions}
\label{sec:conclusions}

We have outlined a method for kinematic bulge-disk decomposition using self-consistent, DF-based dynamical models. The method can be used to model any combination of data, including deep optimal images and 1D/2D kinematic constraints. Our {\galactics}-based implementation (\magrite) is efficient enough ($\sim$1-2 minutes per model on modern CPUs) to fit deep KiDS images and SAMI kinematic maps (\figref{bestfit_samikids_1D}), exactly as conceptualized in \figref{dynamics}. 

We have fit a well-resolved SAMI/GAMA galaxy G79635, showing that the best-fit parameters and posteriors are largely consistent with \profit 2D decompositions and with the independent \HI $W_{50}$ constraint. This suggests that \magrite can extrapolate reasonable rotation curves even without IFS data reaching the peak of a galaxy's rotation curve - a crucial requirement for accurate stellar mass and angular momentum estimates. In the provided appendices, we demonstrate that \magrite can fit synthetic model data with minimal biases. However, we caution that fits to real data are not immune to biases, particularly in the presence of significant non-axisymmetric features. Furthermore, we showed that poorly-fitting models can seriously underestimate parameter uncertainties by yielding artificially narrow posterior PDFs. This can be mitigated but not corrected by estimating uncertainties from mock observations of the best-fit model.

Our example galaxy was selected as a well-resolved, fairly passive spiral galaxy with an HI detection, but there are many more SAMI galaxies with similar quality data. The KiDS images are good enough to constrain the bulge fraction in G79635 to at most a few percent and clearly show deviations from our idealized models, which assume axisymmetric, exponential disks and simple star formation histories for each component. We therefore demonstrate that data quality is not the main impediment to improved, physical modelling of galaxies, but the models themselves. G79635's azimuthally-averaged disk profile can be reproduced with a combination of an unusually thick and smoothly-truncated exponential disk, but would be better fit with a \sersic or non-parametric profile disk including perturbations from spiral arms.

One shortcoming of the method using kinematic moment maps is that these must be derived independently; nonetheless, the method can be generalized to fit spectral datacubes directly \citep[e.g.][]{TabMerAra17} and we plan to implement this functionality within \magrite. Additional model features like spiral arms, dust attenuation/scattering \citep{PasPopTuf13b} and more flexible/non-parametric density profiles are longer-term ambitions. \magrite is under active development and will be released in the near future, alongside early results from a larger SAMI sample. Parties interested in testing the code or contributing to future development are encouraged to contact the authors.

\section{Acknowledgements}
\label{sec:acknowledgments}

This research was conducted by the Australian Research Council Centre of Excellence for All-sky Astrophysics (CAASTRO), through project number CE110001020. This work was supported by the Flagship Allocation Scheme of the NCI National Facility at the ANU. The SAMI Galaxy Survey is based on observations made at the Anglo-Australian Telescope. The Sydney-AAO Multi-object Integral field spectrograph (SAMI) was developed jointly by the University of Sydney and the Australian Astronomical Observatory. The SAMI input catalogue is based on data taken from the Sloan Digital Sky Survey, the GAMA Survey and the VST-ATLAS Survey. The SAMI Galaxy Survey is funded by CAASTRO and other participating institutions. The SAMI Galaxy Survey website is \url{http://sami-survey.org/}. DST acknowledges support from a 2016 University of Western Australia Research Collaboration Award. BC acknowledges support from the Australian Research Council's Future Fellowship (FT120100660) funding scheme. 

\appendix
\section{Integration Scheme Comparison}
\label{app:integration}

\begin{figure*}
\includegraphics[width=\textwidth]{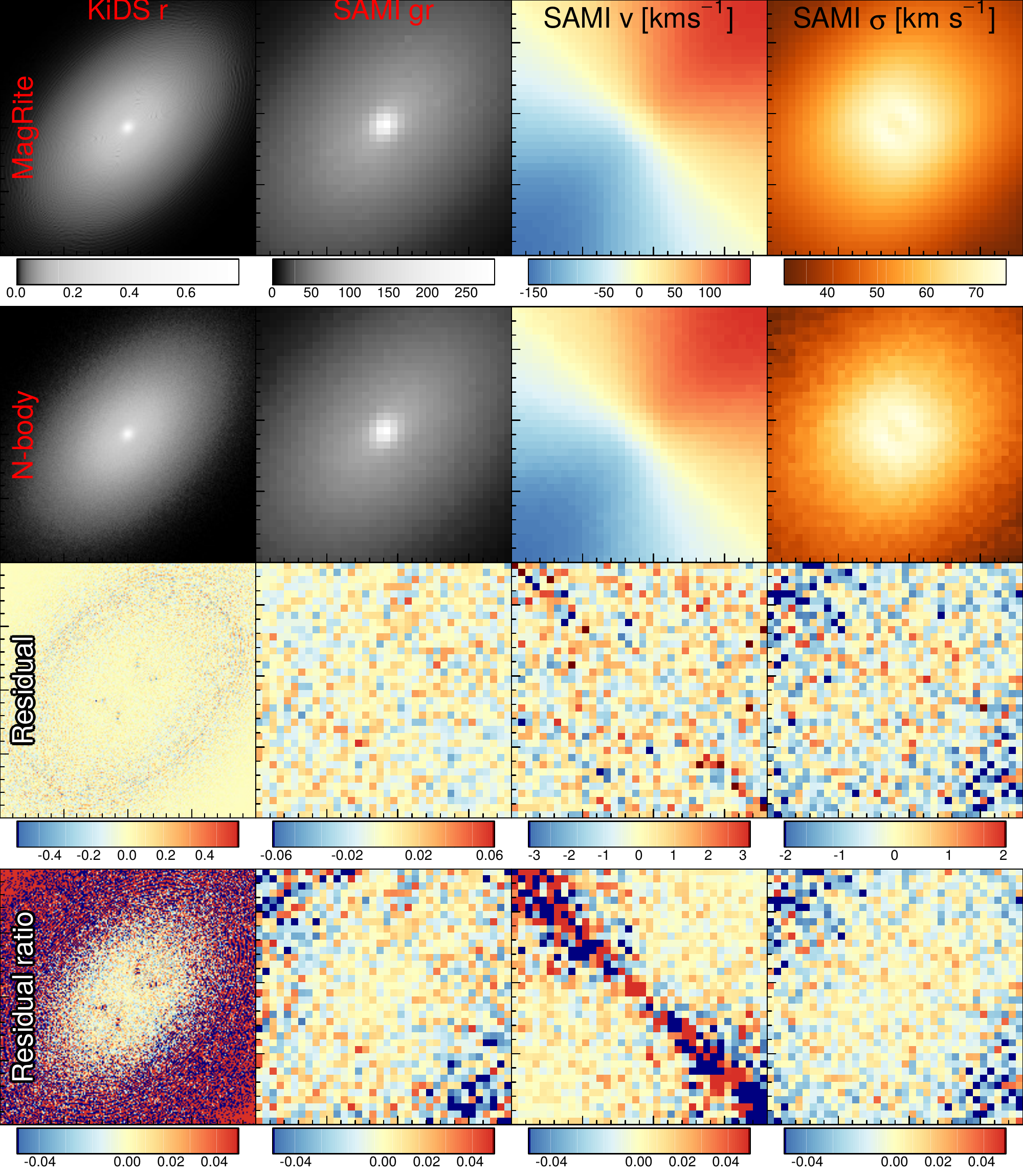}
\caption{Comparison between maps from grid-based and Monte Carlo integration using 20M/0.4M disk/bulge particles, respectively. The residuals are shown on an absolute scale (MagRite - N-body) and as a ratio relative to the smoother MagRite map. The large relative velocity residuals along the minor axis are due to the small absolute value of the velocity; absolute differences are small.
\label{fig:integration_comp}}
\end{figure*}

To test the accuracy and speed of the DF integration scheme described in \subsecref{methods_dfs} and \subsecref{methods_synthetic}, we generate synthetic maps using our method and also from a high-resolution GalactICS model with 20M/0.4M disk/bulge particles, respectively. \figref{integration_comp} shows maps and residuals generated using the best-fit model parameters for G79635 with both of these integration schemes. Despite the large number of particles, the MC (N-body) maps are still shot-noise limited near the outskirts of the disk. Furthermore, sampling this many particles takes nearly 10 minutes on freq a modern test machine (Intel i5-4690 at 3.50GHz). Each accepted particle requires just over three proposal on average, meaning that nearly 70\% of the computing time is effectively wasted evaluating rejected proposals. By contrast, the grid-based integration method takes under a minute and generates smoother, noise-free maps which are virtually indistinguishable from the unbiased N-body maps near the well-sampled galaxy center. This is accomplished mainly by making fewer calls to the expensive DF evaluation methods, effectively spawning dozens to hundreds of particles per DF sample.

As discussed in \subsecref{methods_synthetic}, our grid-based integration scheme is not entirely ideal. Placing evenly-distributed samples at the center of each bin is computationally efficient, but requires a random angular offset in $\theta$ to generate smooth maps - otherwise, the model images would have bright ``spokes'' at the sampled angles $\theta$ and artificial gaps between them. Similarly, binning evenly-spaced ellipses onto a rectangular grid creates Moir\'{e}-like artifacts, apparent as an X-shaped residual in \figref{integration_comp}. These issues could be resolved by distributing DF samples over the projected areas of elliptical rings, rather than as points binned onto a rectangular grid. In principle, this would also need to be done in 3D, i.e. by generating rings corresponding to the top of one bin and the bottom of the next. These would be fairly inexpensive calculations compared to the other steps in model integration, but are somewhat complex and would be unlikely to change the PSF-convolved model maps significantly, so they are left to future revisions of \magrite.

One measurable impact of the integration scheme is the change in model likelihoods through stochasticity. To test this, we generate a series of maps for the best-fit model, varying only the random seed used to generate the angular offsets in $\theta$. For the $r$-band KiDS image, the $\chi^2$ value varies by about 40 from different seeds. This is an insignificant difference for a well-fitting model but highly significant for one with a large $\chi^2_{red}$, as discussed in \appref{stats}. To minimize the impact of this issue, we keep the random seed fixed for all fits.

\section{Model Stability Test}
\label{app:stability}

\begin{figure*}
\includegraphics[width=\textwidth]{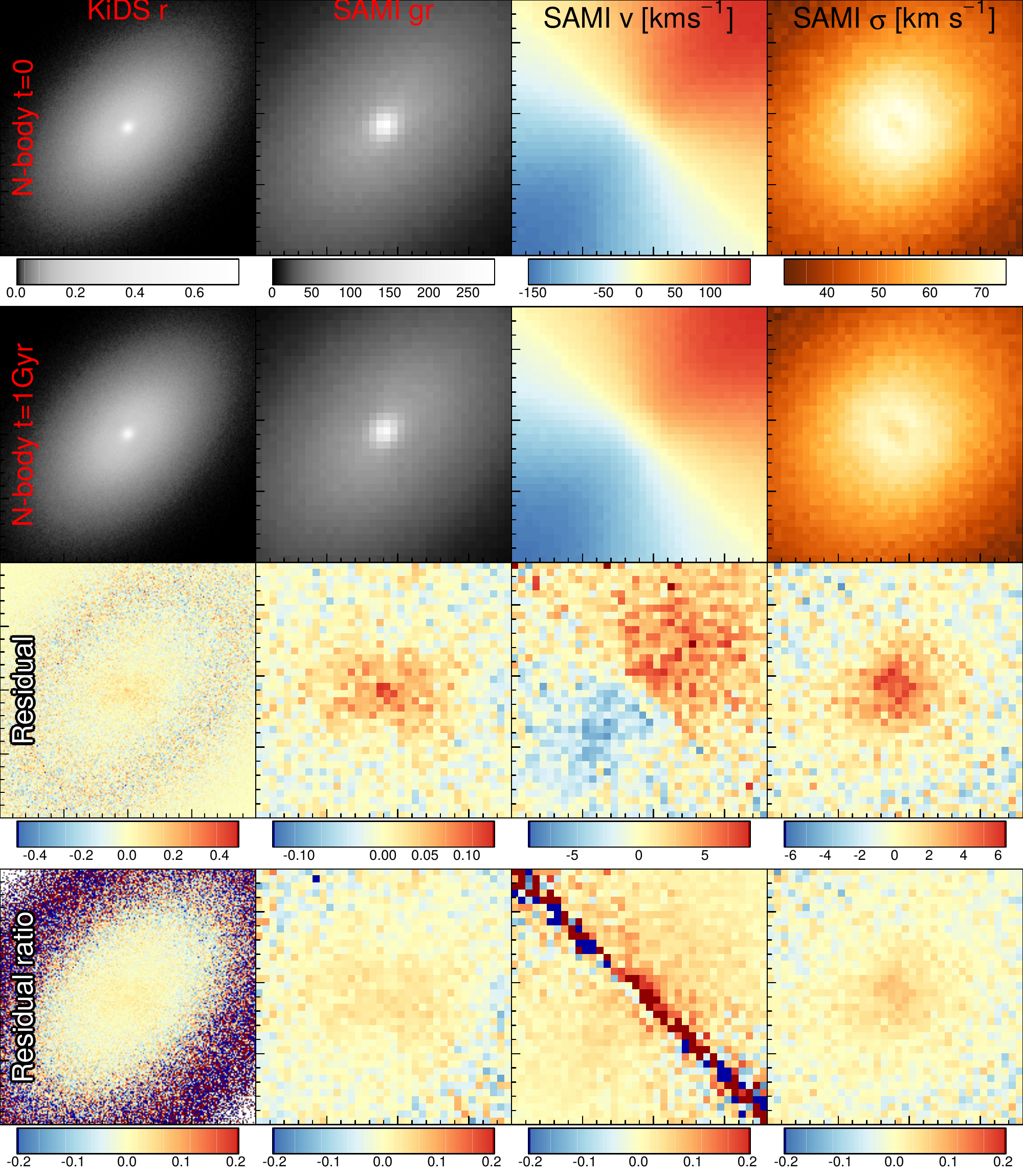}
\caption{Comparison between maps generated from N-body initial conditions and after 1~Gyr of evolution. Pixels outside of the limits of the color bar are assigned the darkest (most saturated) colors. The residuals show some evolution in the galaxy structure due to the thick, truncated disk being slightly out of equilibrium.
\label{fig:nbody_evol}}
\end{figure*}

To test the long-term stability of the model, we generated N-body initial conditions with \galactics, sampling the disk/bulge/halo with 5M/0.1M/5M particles and using softening lengths of 50/50/150 pc, respectively. We ran the model for 1~Gyr with PARTREE \citep{Dub96}, using a fixed 0.196Myr timestep and opening angle of 0.8. \figref{nbody_evol} shows the resulting maps from the evolved galaxy compared to the initial conditions. There is evidence of relaxation of the system, with the evolved model having lower central density and velocity dispersion and a shallower rotation curve. 

The evolution of this model is not representative of typical \galactics model. As discussed in \secref{results}, the best-fit model has an unusually large disk scale height and small truncation radius. The initial virial ratio $q=-2T/W$, where $T$ and $W$ are the total kinetic and potential energy, respectively, is $q=1.00275$. While the deviation from unity is not large, the total energy of the system is dominated by the dark matter halo (with nearly 90\% of the mass), so the stellar component is likely out of equilibrium by a few percent. Accordingly, the virial ratio drops below unity by a similar factor and shows damped oscillations before reaching a new equilibrium.

We adjust the model to have a thinner disk ($z_{d}=1kpc$) and much larger truncation radius ($r_{t,d}=8r_{d}$) with the same disk mass. This model shows virtually no evolution outside of the inner 200 pc, where there is a modest depression in the central density and velocity dispersion. We conclude that while the disk truncation parameters and scale height can in principle mimic a non-exponential ($n_{s}<1$) disk, adjusting them beyond \galactics limits can produce unstable models and should be avoided. This could be accomplished without running simulations simply by placing stronger priors model parameters or on output diagnostics like the virial ratio, but further testing is needed to determine guidelines for these limits.


\section{Model Fitting Test}
\label{app:fitting}

We test the code's ability to recover model parameters by fitting synthetic maps generated by \magrite, a process that is . We assume shot noise-dominated errors for the flux maps, given a gain and mean sky brightness in counts per pixel. The higher-order SAMI moment maps require some simplifying assumptions. Kinematic constraints originate mainly from stellar absorption lines, which only cover a small fraction of typical spectra. We parameterize this effect with a simple ``kinematic gain'' ratio $g_{k,eff}$, which is roughly the ratio of the sum of the equivalent widths of all absorption lines to the full wavelength range. We generate a noisy flux-weighted $v_{LOS}$ DF for each spaxel, where the counts in each bin are multiplied by $g_{k,eff}$, and fit a Gaussian to extract the mean velocity and dispersion. Finally, we re-use the existing masks and also the velocity and dispersion error maps for consistency with the original fits. We also adjust $g_{k,eff}$ until the noise in the velocity/dispersion maps is roughly consistent with the original errors.

\begin{figure*}
\includegraphics[width=\textwidth]{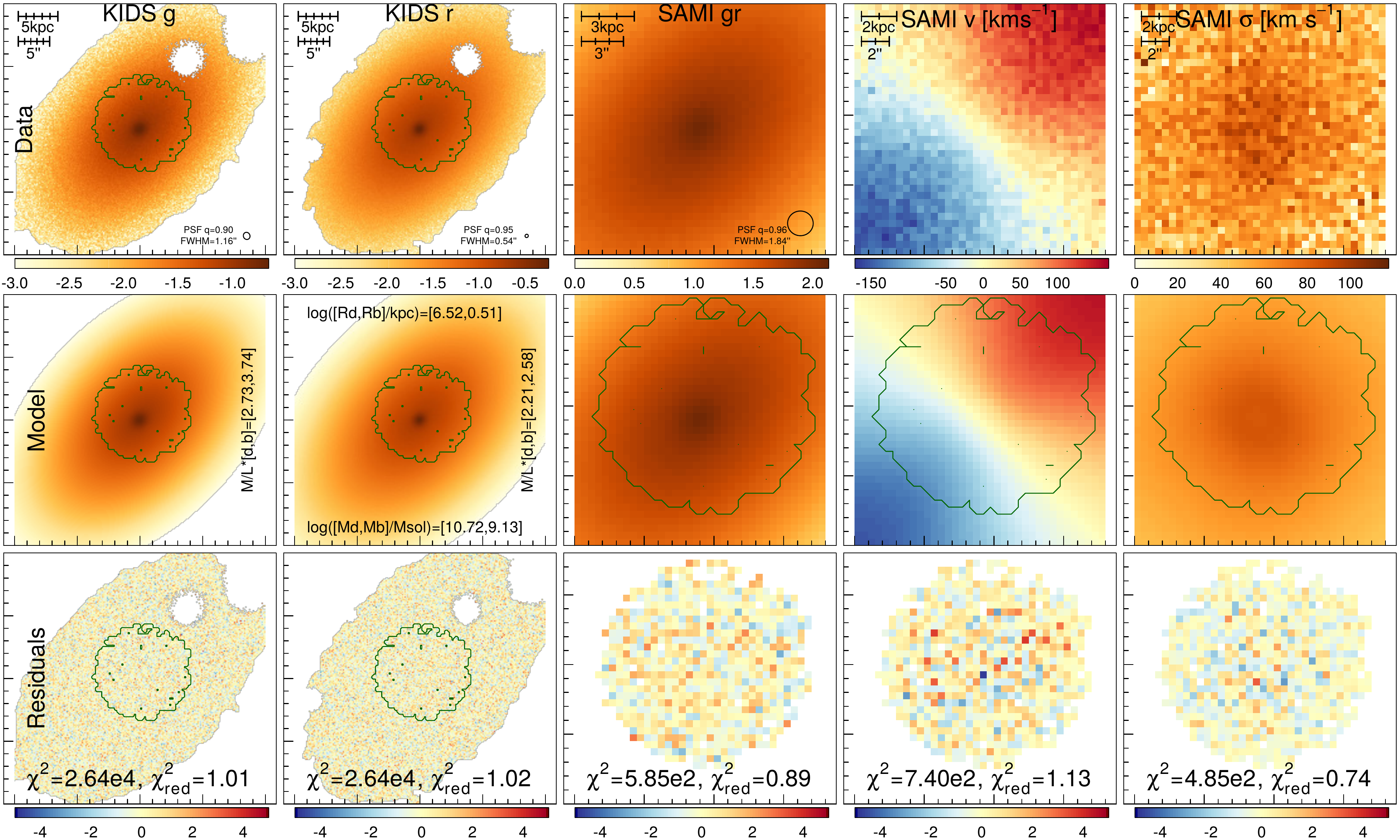}
\caption{Mock G79635 maps with realistic shot noise. The kinematic maps are derived from Gaussian fits to noisy $v_{los}$ DFs and therefore do not necessarily follow Poisson or (approximately) Normal statistics.
\label{fig:bestfit_mock}}
\end{figure*}

\figref{bestfit_mock} shows synthetic noisy maps for G79635 using $g_{k,eff}=0.02$. As expected, the flux maps are completely consistent with (nearly) Normal shot noise and have $\chi^2_{red}\approx1$. However, the noise in the $v_{LOS}$ and $\sigma$ maps is not entirely identical to that from the original SAMI maps, being slightly over- and under-estimated, respectively. This not surprising, given the non-linear nature of kinematic fits, but the mock kinematics still appear consistent with random noise without any obvious systematic bias and are usable as data for a mock fit. The possibly conservative choice of $g_{k,eff}=0.02$ is motivated by the need to keep this noise consistent with the SAMI error maps; larger values lead to excessively smooth kinematic maps. When fitting, we continue to use directly-measured velocity moments in the model, rather than the Gaussian fits used for both the mock and real data. This is because fitting $v_{LOS}$ requires some estimate of the uncertainty of the $v_{LOS}$ DF in each spaxel, and it is not clear what this uncertainty should be in a noiseless model. Since G79635 has a very modest bulge, there is little difference between the two measurements; however, this may not be the case in galaxies with more massive and extended bulges, where this issue would be worth revisiting.

\begin{figure*}
\includegraphics[width=\textwidth]{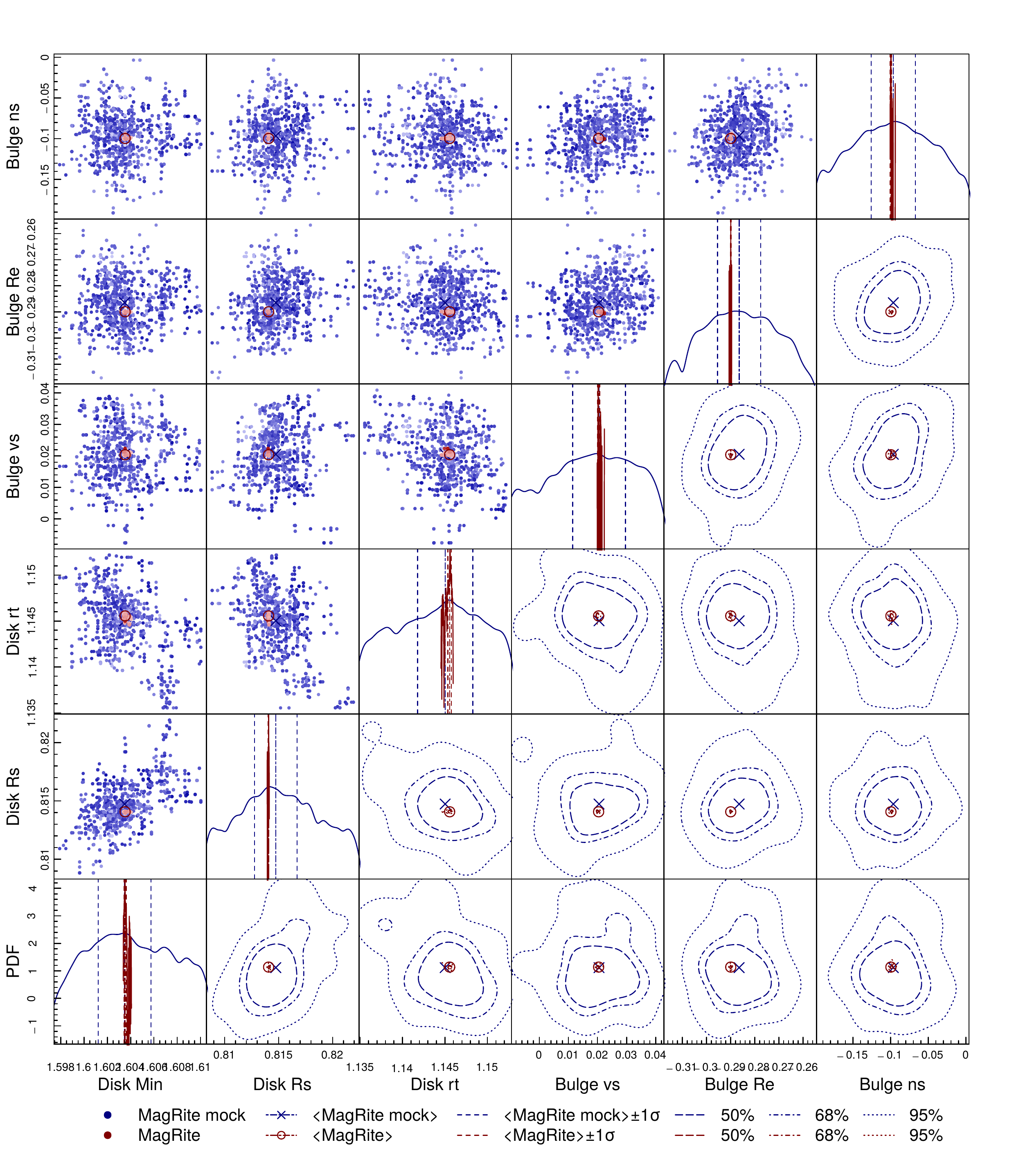}
\caption{Triangle plot showing joint posterior distributions for model input parameters (Disk Min$\equiv\log10{(M_{d,in}/M_{sim})}$, Disk Rs$\equiv\log10{(R_{d}/\mathrm{kpc})}$, Disk rt$\equiv\log10{(r_{t,d}/\mathrm{kpc})}$, Bulge vs$\equiv\log10{(v_{b}/\mathrm{v_{sim}})}$, Bulge Re$\equiv\log10{(R_{b}/\mathrm{kpc})}$, ns$\equiv\log10{(n_{s})}$). Points are color-coded by log probability as described in \figref{mcmc_samikids}. PDFs are shown for fits to the mock data and the observed data. Since the latter are exceptionally narrow, univariate PDFs along the diagonal are shown on a logarithmic scale spanning six orders of magnitude.
\label{fig:mcmc_mock}}
\end{figure*}

\figref{mcmc_mock} shows posterior distributions for a \magrite fit to the mock data shown in \figref{bestfit_mock}. Reassuringly, the best-fit parameters are close to the inputs, with small deviations well within the $1\sigma$ uncertainties. Small offsets are expected given the noise and the difference between directly measured and fitted kinematics, but there are no significant systematic biases. This is not to say that fits to real data will be devoid of biases - this simply verifies that \magrite recovers unbiased parameters when the data are perfectly described by the model.

One possible concern is that there is still some noise in the model probabilities themselves - when colored by log probability, the points do not show smooth gradients in posterior probability from the maximum likelihood solution. This is especially true for the fits to the observed data, which is unfortunately not clearly visible in \figref{mcmc_mock} because the posteriors are so narrow. This is likely due to the issues with model integration outlined in \appref{integration}. The main practical effect of a slightly stochastic model likelihood is that convergence to the best-fit solution can be slow, as the solution wanders between entirely artificial local maxima. This problem is exacerbated when the best-fit model is a poor fit, since even extremely small changes in input parameters can cause spurious changes in the model likelihood. Fortunately, since the actual changes in parameter values tend to be small, the impact on the posteriors from mock fits is minimal.

Since poorly-fitting models tend to vastly underpredict parameter uncertainties (for reasons detailed in \appref{stats} below), we suggest that uncertainties from mock fits should be considered lower bounds on the ``true'' parameter uncertainties. Again, this does not guarantee that there are no biases in the best-fit parameters themselves, so caution must be taken in defining priors and interpreting best-fit values in such cases. Similarly, since the mock data are idealized, they should only be interpreted as lower limits until further testing is done to quantify the impact of deviations from the assumed model parameterizations, which are necessarily present in all galaxies.

\section{Parameter Uncertainties from Poorly-Fitting Models}
\label{app:stats}

It is not necessarily intuitive that poorly-fitting models can or should yield systematically smaller uncertainties than good models. Nonetheless, it is a natural consequence of the shape of some common statistical distributions. The difference in log-likelihood ($\Delta LL$) between an $n$- and $(n+1)$-$\sigma$ deviation from a univariate Normal distribution is $-(n+0.5)$. The difference between 10- and 11-$\sigma$ deviations is then much more significant than that between 2- and 3-$\sigma$ deviations, simply because the log of the PDF of a Normal distribution declines as $x^{-2}/2\sigma$.

For a more practical example, the KiDS $r$-band image for G79635 has 25961 usable data points (unmasked pixels). Using the chi-square distribution with 25961 degrees of freedom as the fit statistic\footnote{We neglect model parameters in the effective degrees of freedom, as the number of model parameters in non-linear models is poorly defined \citep{AndSchMel10}.}, the maximum likelihood solution for an ideal model has $\chi^2$=25959. A three-$\sigma$ deviation from a univariate Normal distribution (i.e. for a Gaussian parameter posterior) has a log-likelihood 4.5 lower than the peak. The equivalent range of likelihoods for the given $\chi^2$ distribution is $\chi^2=[25281.4,26648.6]$, i.e. $\Delta\chi^2=[-677.6,-689.6]$ or $\chi^2_{red}={0.974,1.026}$. However, the best-fit model was only able to achieve $\chi^2=1.677e3$ for the $r$-band image, or $\chi^2_{red}=6.46$. An increase in $\chi^2$ of just 10.6 produces a $\Delta LL=-4.5$, so the range of acceptable $\chi^2$ shrinks by a factor of approximately $\chi^2_{red}$. Accordingly, the posterior parameter PDFs shrink by a comparable margin, depending on the linearity of the model. For a large number of degrees of freedom, this behaviour approaches that of a Normal distribution.

Finally, we note that overfitting is strongly disfavoured by the $\chi^2$ statistic - a desirable feature for data with reliable errors, as in images dominated by shot noise from a large number of counts. For lower signal-to-noise and/or where other terms like read noise are important, Poisson statistics or the so-called \citet{Cas79} statistic should be used.

\section{Thick Disk Surface Brightness Profile Fits}
\label{app:thickdisk}

\begin{figure*}
\includegraphics[width=\textwidth]{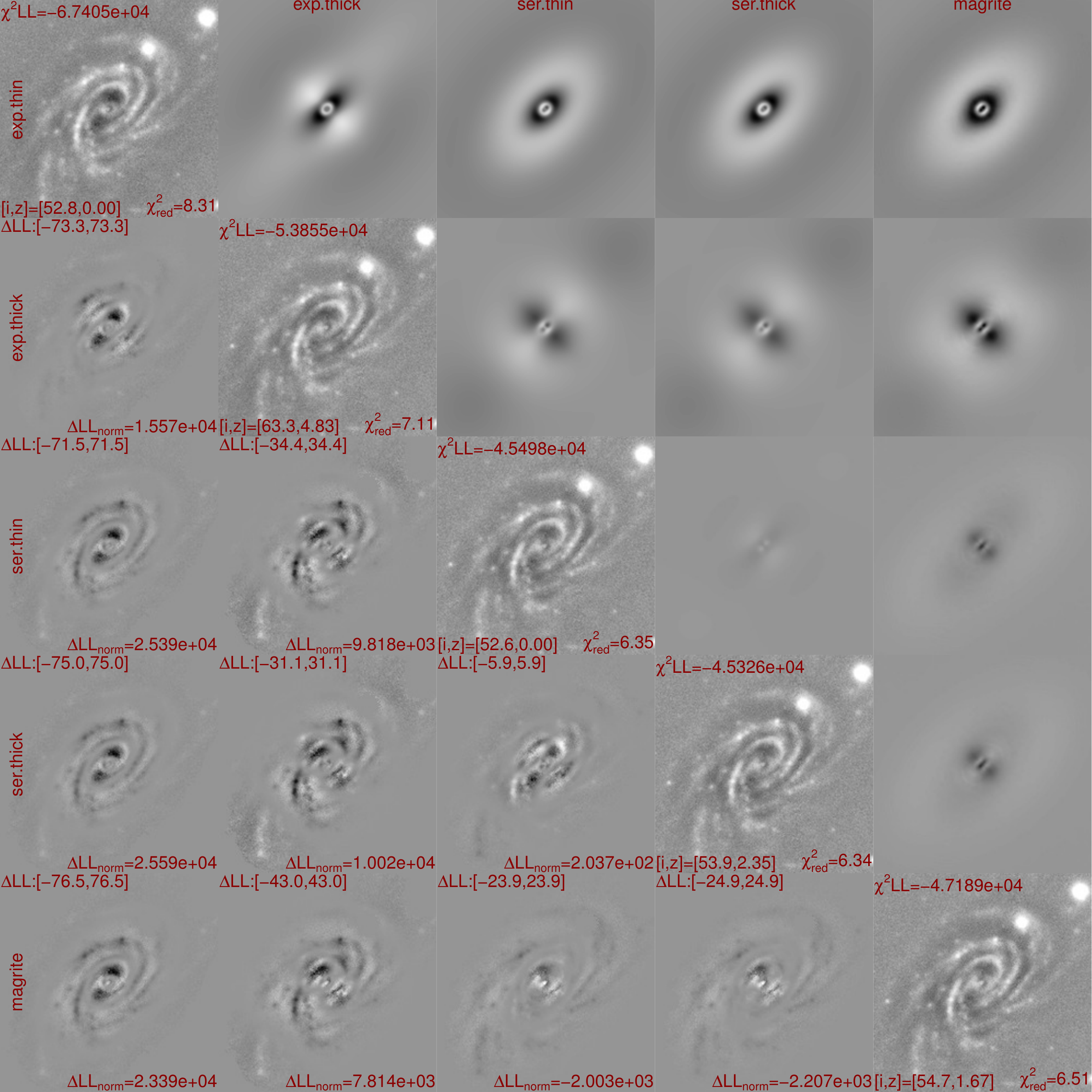}
\caption{Comparison of $r$-band residuals for five G79635 model fits. The models include four ProFit models - thin exponential disk (``exp.thin''), thick exponential disk (``exp.thick''), thin \sersic disk (``ser.thin'') and thick \sersic disk (``ser.thick'') - as well as the \magrite fit. Diagonals panels show model residuals $\chi=(data-model)/\sigma$ on a common scale (from approximately $-22 < \chi < 22$), where dark colours correspond to excess in the model and light colours to excess in the data. Also listed are the log-likelihood assuming a $\chi^2$ distribution; $\chi^2_{red}$; and the disk inclination $i$ and scale height in kpc $z$. Panels above the diagonal show differences between models on a common linear scale centered on zero; dark colours are where the model in a given row is brighter than the model in the given column. Panels below the diagonal show differences in model residuals ($\chi$) on arbitrary linear scales, again centered on zero and where dark colours represent a better fit in the the model in the given row than in the model in the given column. The range of changes in per-pixel log-likelihood (assuming Normally distributed errors) is show in the top left, and the total change in Normal log-likelihood is listed at bottom-right. As expected, residuals improve with added model complexity, but are slightly worse in \magrite (which also fits the kinematics and $g$-band image) than in the \profit\xspace\sersic thick disk fit; however, both \profit thick disk models yield excessively large scale heights.
\label{fig:thick}}
\end{figure*}

As discussed in \secref{results}, the best-fit \magrite model has an unusually large disk scale height of 1.67 kpc. We have run a number of tests by modifying \profit to fit a thick disk with a $\mathrm{sech^2}$ vertical profile using a similar integration scheme as described in \subsecref{methods_dfs}. We superpose thirty \sersic disks above and below the disk midplane by shifting the profile center along the minor axis, weighting each disk by the total mass within each vertical bin. This is not the most efficient integration method for a 3D density profile and has limited accuracy for highly inclined thick disks, but it is analogous to the \magrite method and ideal for model comparisons.

The results of fitting thick disk profiles to G79635's $r$-band image are shown in \figref{thick}. Firstly, fitting a thick exponential disk (second row/column) significantly improves the residuals over a thin exponential. The improvement is seen largely in the two under-dense regions between the galaxy center and spiral arms (roughly NNE and SSW from the galaxy center). However, the thin \sersic disk achieves similar improvements without significantly worsening the residuals anywhere else in the disk. Thus, it is clear that a thick disk can compensate for deviations from a pure exponential disk, but not necessarily as well as by simply modifying the radial profile. Unfortunately, the best-fit \profit disk thicknesses are unrealistically large - 4.83 kpc for the exponential disk and 2.35 for the \sersic disk.

\begin{figure*}
\includegraphics[width=\textwidth]{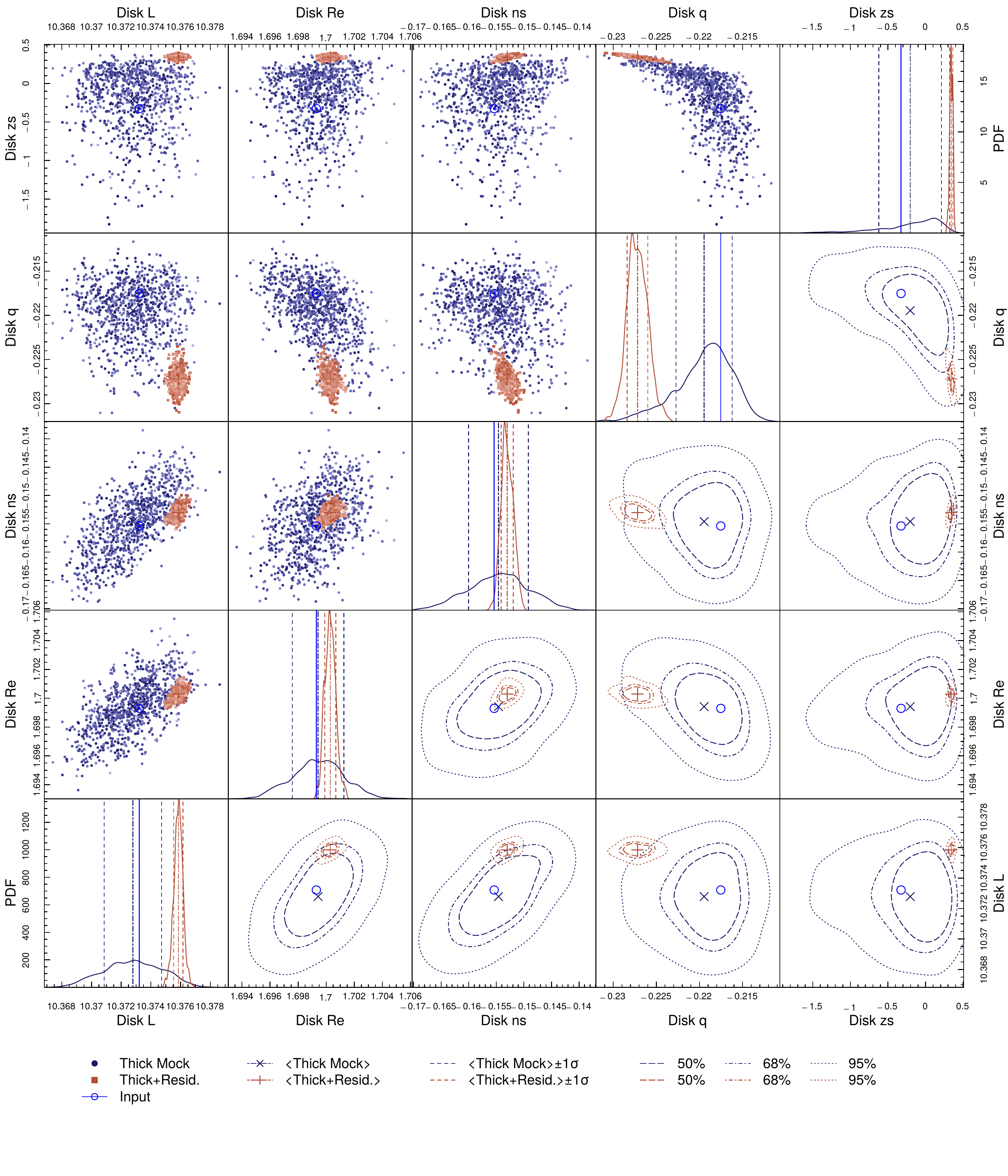}
\caption{Triangle plot showing joint posterior parameter distributions ($L\equiv\log10{(L_{r}/\lsole)}$, $Re\equiv\log10{(R_{e}/\mathrm{kpc})}$, $n\equiv\log10{(n_{s})}$, $zs\equiv\log10{(z_{d}/kpc)}$, and where $q$ is the disk axis ratio) for \profit $r$-band \sersic disk fits to mock data, with and without including clipped residuals from the best-fit model. Panels are structured as in \figref{mcmc_samikids}. The fit to the mock data with clipped residuals (``Thick+Resid.'') has smaller uncertainties and is also significantly biased, particularly for the scale height and inclination. The magnitude bias is due to the small net positive flux of the residuals. For clarity, points in the upper-left quadrant are thinned by a factor of 5 and 10 for the ``Thick Mock'' and ``Thick+Resid.'' samples, respectively.
\label{fig:thickmcmc}}
\end{figure*}

\figref{thickmcmc} shows posterior distribution from \profit $r$-band fits to mock data using the same input parameters. The ``Thick'' model shows the same chains as in \figref{mcmc_samikids}, where the mock data was generated with a more plausible scale height $z_{d}=0.1R_{e,d}/1.67835$ (equivalent to $0.1R_{d}$ for an exponential disk), and with the other best-fit parameters taken from the best thin disk fit. A second fit was run on mock data with clipped residuals added back in. Specifically, after generating the PSF-convolved model image, we add $\chi\sigma\mathrm{tanh(abs}(\chi/1.25))^{30}$, where $\sigma$ is the per-pixel uncertainty. The tanh scaling smoothly truncates residuals below $2\sigma$, so the disk more closely follows a \sersic profile. Because overdensities like spiral arms and star-forming regions tend to be more significant than underdensities, the residuals have a small net positive flux of slightly under one percent of the disk luminosity, which is reflected in a small positive bias in disk luminosity in \figref{thickmcmc} compared to the input parameters. The size and \sersic index are somewhat biased, but the scale height and axis ratio are significantly over- and under-estimated, respectively, indicating that structured residuals with a small net flux can severely bias poorly-constrained and/or degenerate parameters even when the model is a reasonable approximation to the data.

The fact that scale height and inclination are highly degenerate is not surprising - if the disk's vertical density profile is the same as its radial profile, then the scale height and inclination will be completely degenerate. Using a $\mathrm{sech}^2$ vertical density profile rather than exponential limits but does not prevent this degeneracy. \magrite achieves a tighter constraint on the inclination by fitting the velocity map. Of course, the mass model also modifies the rotation curve, but the stellar mass is independently constrained by the flux maps. Unfortunately, the kinematic constraints on the disk scale height itself are weak. In principle, the disk dispersion is related to the disk's vertical structure, but this is (mostly) independently parameterized in \galactics by $\sigma_{R0}$, and SAMI's spectral resolution is not fine enough to measure typical disk dispersions anyway. Thus, there is insufficient data to guarantee an accurate best-fit scale height, and a strong prior based on observations of edge-on disks should be used in practice.

\section{Summary of the GalactICS Method}
\label{app:galactics}

We provide a brief description of the methods used by \galactics to generate DFs for composite galaxy models containing any number of disk-like and spherical components composed of stars, dark matter and gas \citep{KuiDub95,WidDub05,WidPymDub08}. We use the \citet{BinTre08} convention for defining the cylindrical radius as uppercase $R$ and spherical radius as lowercase $r$ below, as well as the physics/ISO 80000 convention of $\theta$ for the polar angle and $\psi$ for the azimuthal angle (in contrast with \subsecref{methods_dfs}.  We also use the relative binding energy $\ec\equiv E$ and relative potential $\Psi\equiv -\Phi$ where $\ec = \Psi - v^2/2$ and $v$ is the velocity (such that $v^2/2$ is the specific kinetic energy).

A galaxy model is defined parametrically by mass profiles for each collisionless component.  Stellar bulges and dark matter halos are defined by spherical models described by a radial profile $\rho(r)$. 
For example, we use the 3D deprojected Sersic profile \citep{PruSim97} to describe the bulge, given by:
\begin{equation}
\rho_b(r) = \rho_0 \left ( \frac{r}{R_e} \right )^{-p} \exp[-b_n(r/R_e)^{1/n_s}],
\end{equation}
where the parameter are a characteristic density $\rho_0$, the projected half-mass radius $R_e$, and the the Sersic index $n_s$. The two parameters $b_n$ and $p$ are structural quantities depending on $n_s$.
They are well-approximated by the formulae:
\begin{eqnarray}
p &=& 1 - 0.6097/n_s + 0.05563/n_s^2 \\
b &=& 2 n_s - 1/3 + 0.009876/n_s
\end{eqnarray}
for $0.6<n_s<10$ and $10^{-2} < R/R_e < 10^3$ \citep{TerGra05}.

There are numerous options for halo profiles depending on one's theoretical bias, including profiles with constant density cores or power-law cusps \citep{MerGraMoo06}.  In this paper, we use a double power-law model to describe the halo:
\begin{equation}
\rho_h(r) = \rho_s \left ( \frac{r}{r_s} \right )^{-\alpha} \left( 1 +
\frac{r}{r_s} \right)^{\alpha-\beta}
\end{equation}
where $(\alpha,\beta)=(1,3)$ is the NFW profile and $(\alpha,\beta)=(1,4)$ is the \citet{Her90} profile; such double power-law models are sometimes referred to as generalized NFW or Hernquist profiles.

Disks are flat axisymmetric models and are approximated by the density law:
\begin{equation}
\rho_d(R,z) = \frac{M_d}{4\pi R_d^2 z_d} \exp(-R/R_d) {\rm sech}^2(z/z_d)
\label{eq-disk}
\end{equation}
\citep{vdKSea81}.
We emphasize that this is not the exact disk density but rather a close approximation to the final density law derived in the computation of the disk distribution function defined below.

Finally, we force the density of each component to smoothly approach zero by multiplying each profile by a truncation function.  We truncate density laws using a logistic function defined by:
\begin{equation}
T(t) = (1 + e^t)^{-1},
\label{eq-logistic}
\end{equation}
with 
\begin{equation}
t = \frac{r - r_t}{\delta r_t},
\end{equation}
where $r_t$ is the truncation radius and $\delta r_t$ is the radial width of the truncation interval. Equation \ref{eq-logistic} is a simple representation of a smooth step function chosen for its computational efficiency and continuous derivatives.

The method computes an axisymmetric DF for the system of the form:
\begin{equation}
f(\ec,L_z,\ec_z) = f_d(\ec,L_z,\ec_z) + f_b(\ec) + f_h(\ec)
\end{equation}
where $\ec$ is the relative binding energy $\ec\equiv -E$, $L_z$ is the $z$-component of angular momentum and $\ec_z$ is the $z$ energy defined below.  The bulge and halo are functions of energy alone and so are modelled as spherical isotropic systems.  The disk DF is defined as a function of three integrals of motion.  The first two are the usual energy and $z$-component of angular momentum for axisymmetric systems but we introduce a third approximate integral $\ec_z=\Psi_z - v_z^2/2$ where $\Psi_z$ is the vertical potential defined as $\Psi_z\equiv \Psi(R,z) - \Psi(R,z=0)$, where $\Psi$ is the relative gravitational potential of the system $\Psi = -\Phi$.

With these various definitions in hand, we can describe a numerical procedure for computing the component DFs.  First consider a purely spherical system composed of multiple components -- we will consider modifications when including a thin disk component later. The construction of an isotropic DF $f(\ec)$ from a potential-density pair can be accomplished using Eddington's formula \citep[e.g.,][]{BinTre08}
\begin{equation}
f(\ec)=\frac{1}{\sqrt{8}\pi^2}\left[ \int_0^\ec
\frac{d\Psi}{\sqrt{\ec-\Psi}}\frac{d^2\rho}{d\Psi^2} +
\frac{1}{\sqrt{\ec}}\left(\frac{d\rho}{d\Psi}\right)_{\Psi=0} \right].
\end{equation}
For a system of total density $\rho(r)$, we can compute the total potential using the integral expression:
\begin{equation}
\Psi(r) = 4\pi G \left[\frac{1}{r}\int_0^r dr' r'^2\rho(r') + \int_r^\infty dr' r'
\rho(r')\right]
\end{equation}
To use the Eddington formula, one needs to determine the function $\rho(\Psi)$ and its derivatives up to second order.  In general, it is difficult to find an analytic solution so we use the following numerical method.  We first define a grid with $n$ radial positions equally spaced in logarithmic space defined by:
\begin{equation}
u_i = \log r_i/r_0
\end{equation}
where $r_0$ is a reference radius and $r_i$ is the grid point radius for $i=1..n$.  With this transformation, we can compute the potential at the grid positions $u_i$ as:
\begin{multline}
\Psi(u_i) = 4\pi G r_0^2 \times \\
\left[ e^{-u_i}\int_{-\infty}^{u_i}
du'e^{3u'}\rho(u') + \int_{u_i}^\infty du' e^{2u'}\rho(u')\right].
\end{multline}
In practice, the infinite limits for the inner and outer integrals can be replaced with the initial and final points of the logarithmic grid $u_1$ and $u_n$ without loss of accuracy.  The inner integral is just the mass versus radius and this becomes insignificant if the inner most radius is sufficiently small.   For the outer integral, since the density drops to zero at a finite radius defined by the truncation function there is no need to integrate beyond this point. Accurate and stable numerical solutions are achievable with 200 grid spacings per dex, making the new method much faster.

The integral in Eddington's formula can be solved numerically by creating a tabulated function of density $\rho$ for each of the model functions versus the total $\Psi$ on the logarithmic radial grid.  We use the interpolation modules in the GNU science library (GSL) to create a splined function for $\rho(\Psi)$ and its first and second derivatives.  We can then solve the integral in the Eddington's formula numerically to determine a DF for each spherical component independently.  The DF is also determined as tabulated function by finding $f$ for each value of $\ec = \Psi$ on the radial grid.  

To incorporate the disk component in this scheme, we use the {\em ad hoc} method introduced by \citet{WidDub05}.  The disk density law of equation is spherically averaged to create an additional spherical density component that is part of the total potential.  The individual DF's computed from the scheme above then include a good approximation  of the disk potential in their derivation. We see below that when we include the flattened disk, the spherical density profiles of the bulge and halo that are consistent with their DF's are modified slightly from the ideal spherical case but remain close to the original definition.  

We now consider the construction of the disk DF.  \citet[][,hereafter KD95]{KuiDub95} introduced this DF to describe the disk:

\begin{multline}
f_d(E_p,L_z,E_z) = 
\frac{\Omega(R_c)}{(2\pi)^{3/2}\kappa(R_c)}
\frac{\widetilde{\rho}_d(R_c)}{\widetilde{\sigma}_R^2(R_c)\widetilde{\sigma}_z(R_c)} \times \\ 
{\rm exp}\left[-\frac{E_p-E_c(R_c)}{\widetilde{\sigma}_R^2(R_c)} -
\frac{E_z}{\widetilde{\sigma}_z^2(R_c)}\right]
\label{eq-diskdf}
\end{multline}
where $E_p=E-E_z$ is the energy in planar motions, $L_z$ is the $z$-component of angular momentum, $R_c$ and $E_c$ are the radius and energy of the circular orbit with angular momentum $L_z$, and $\Omega$ and $\kappa$ are the circular orbital and radial frequencies derived from the total potential.  As shown in KD95, one can obtain the density by integrating over velocities and the resulting density in the plane is $\widetilde{\rho}_d(R)$ with fractional errors $O(\widetilde{\sigma}_R^2/v_c^2)$.  

The disk density can be obtained by integrating the disk DF over the velocities and the result generates a midplane disk density equal to $\widetilde{\rho}_d(R)$ plus fractional terms of $O(\widetilde{\sigma}_R^2/v_c^2)$.  By construction, this disk DF works best for cool, thin disks where the epicyclic approximation is valid for disk star orbits though warmer and thicker disks are still good equilibria in practice (see \appref{stability}).

The goal is to find a set of ``tilde'' functions in the disk DF $\widetilde{\rho}$, $\widetilde{\sigma}_R$ and $\widetilde{\sigma_z}$ that closely approximate the disk density in equation \ref{eq-disk}.  To this end, we use the density law:
\begin{multline}
\rho_d(R,z) = \frac{M_d}{4\pi R_d^2 z_d} e^{-R/R_d}\times \\
\exp\left[C_1 \frac{\Psi_z(R,z)}{\Psi_z(R,C_0 z_d)}\right]
T\left(\frac{r-r_t}{\delta r_t}\right).
\label{eq-disk2}
\end{multline}
The constants $C_0$ and $C_1 = \ln {\rm sech}^2(C_0)$ are chosen so that the run of vertical density of the disk at a given radius $R$ approximates ${\rm sech}^2(z/z_d)$.   In practice, a good choice is $C_0=3$ corresponding to equivalence of the vertical density at 3 scale-heights for the target density of equation \ref{eq-disk} and this disk density.

At this point, we have built a DF for each component with the density defined in terms of the total potential $\Psi(R,z)$.  To achieve self-consistency, one needs to solve for the total potential:
\begin{equation}
\nabla^2 \Psi(R,z) = -4\pi G [\rho_d(R,\Psi,\Psi_z) + \rho_b(\Psi) + \rho_h(\Psi)]
\label{eq-poisson}
\end{equation}
As in KD95, we use the iterative method of \citet{PreTom70} to find a numerical solution. The method proceeds by making an initial guess of the potential $\Psi$, computing the densities on the right side of equation \ref{eq-poisson} and then re-solving for $\Psi$.  This process is iterated until $\Psi$ relaxes to a solution. In practice, we use a multipole expansion of $\Psi$ in spherical coordinates $(r,\cos\theta)$ with the radius defined on the logarithmic grid. For an axisymmetric system, we can write the potential on the logarithmic radial grid as the multipole expansion:
\begin{multline}
\Psi(r,\theta) = 4\pi G r_0^2 \sum\limits_{l=0}^\infty \frac{P_l(\cos\theta)}{2l + 1}\times \\
\left(e^{-(l+1)u}\int_{-\infty}^u du\;
e^{(l+3)u} a_l(u) + e^{lu}\int_u^{\infty} du\; e^{(2-l)u}a_l(u)\right)
\end{multline}
where $u=\log(r/r_0)$ and
the functions $a_l(u)$ are given by
\begin{equation}
a_l(u) = \int \sin\theta\; d\theta P_l(\cos\theta) \rho(u,\theta)
\end{equation}
\citep[e.g.,][]{BinTre08}.
We can replace the infinite limits by the end points of the grid as before without losing accuracy.  We use convergence of the ``tidal'' radius of the total density within some error tolerance to stop the iteration.  The tidal radius is just the finite radius where the total density drops to zero for the model. The series is truncated for some $l$ sufficiently large to approximate the flattened potential.  

We describe some modifications to this procedure that speed up convergence and overall accuracy.  The initial guess to the total potential is the composite spherical potential derived in the first stage. We first perform an iterative sequence to monopole order until convergence. In a second phase, we restart the iteration with this solution gradually adding the terms for higher order expansions.  Furthermore, the new potential derived at each iteration is determined as a weighted average of the newly determined potential and the previous one. In general, convergence to a high order in $l$ can be reached in a few tens of iterations.  

Following KD95, we also use an analytic ``high harmonics" disk potential to improve the accuracy of the multipole expansion at lower order in $l$. This analytic potential is:
\begin{equation}
\Psi_d^\dagger = \frac{G M_d z_d}{2 R_d^2} \ln{\rm cosh}(z/z_d) e^{-r/R_d} T\left(\frac{r-r_t}{\delta r_t}\right)
\end{equation}
where one notes that the radial parameter is the spherical radius.  One can
use the identity:
\begin{multline}
\nabla^2 f(r) \ln{\rm cosh}(z) = f''(r)\ln{\rm cosh}(z) + \\
\frac{2 f'(r)}{r}[z{\rm tanh}(z) + \ln{\rm cosh}(z)] + f(r){\rm
sech}^2 (z)
\end{multline}
to derive an analytic disk density $\rho_d^\dagger$ from $\Psi_d^\dagger$. The last term reproduces a ${\rm sech}^2(z)$ disk to $O(z/R)^2$ while the other terms are generally small for thin disks and tend to zero at the origin.  In the iterative method described above, we replace the disk density with the residual $\delta\rho_d = \rho_d - \rho_d^\dagger$ when determining the total potential from the multipole expansion.  The total final potential is computed as the sum of the multipole expansion plus $\Psi_d^\dagger$.  In practice, this method allows an accurate representation of the thin disk model for expansion orders of $l<10$ greatly speeding up the overall procedure.

The final step involves the finding ``tilde" functions in the disk DF that match the disk density in equation \ref{eq-disk2}. The first function $\widetilde{\sigma}_R^2(R)$ can be set arbitrarily but following KD95 we use the observationally inspired profile $\widetilde{\sigma}_R^2 = \sigma_{R,0}^2 \exp(-R/R_d)$ where the central radial velocity dispersion $\sigma_{R,0}^2$ is a free parameter for the disk.  Normally, this parameter is chosen so that the disk is stable i.e. Toomre $Q>1$ across the radial extent of the disk.  The remaining functions -- the midplane density $\widetilde{\rho}_d(R)$ and the vertical velocity dispersion $\widetilde{\sigma}_z^2$  -- are iteratively adjusted for each radius on the grid such that the midplane density and the density at $z=z_d$ are the same as that in equation \ref{eq-disk2}.   Finally, the tilde functions can be represented numerically with splines and thus the disk DF is fully specified.  

In summary, the final products of this moderately complex procedure are full specified potentials and densities for each component defined by multipole expansions with modifications for the thin disk to improve accuracy for lower order $l$.  Each component also has a well-defined DF determined by Eddington's method for spherical components and the constructed disk DF from equation \ref{eq-diskdf}.  The multipole coefficients of the potential expansion are tabulated on the logarithmic grid and represented as splined functions in the code, allowing rapid evaluation of the potential and density at any point.  The spherical DFs as $f(\ec)$ and the tilde functions are computed at the predefined grid coordinates. The code takes tens of seconds on current hardware to tabulate these functions - depending on the maximum multipole order - making it practical to use for fitting galaxy observations.

\bibliography{paper}

\end{document}